\newif\iflinenumbers
\linenumbersfalse %

\documentclass[10pt,oneside,notitlepage,abstracton,a4paper]{article}

\usepackage{epsfig,scrpage2}
\usepackage{amsmath}
\usepackage{makecell}
\usepackage{graphicx}
\usepackage{booktabs}
\usepackage{arydshln}
\usepackage{setspace} %
\usepackage[utf8]{inputenc}
\usepackage[T1]{fontenc}

  \date{\normalsize \today}

\usepackage{chngcntr}

\newcommand{\setappendixnumbering}{%
  \renewcommand{\thesection}{\Alph{section}}
  \renewcommand{\theequation}{\thesection.\arabic{equation}}
  \renewcommand{\thefigure}{\thesection.\arabic{figure}}
  \renewcommand{\thetable}{\thesection.\arabic{table}}
  \counterwithin{equation}{section}
  \counterwithin{figure}{section}
  \counterwithin{table}{section}
}

\usepackage[bottom]{footmisc}

\iflinenumbers
  \usepackage{lineno}
  \linenumbers
\fi

\usepackage[affil-it]{authblk}

\usepackage{etoolbox} %
\makeatletter %
\patchcmd{\@maketitle}{\LARGE \@title}{\fontsize{16}{19.2}\selectfont\@title}{}{}
\makeatother
\usepackage{geometry}

 \usepackage{pdfpages}

\newcommand{\imagepath}{.}

\usepackage{xcolor}
\definecolor{DESYcyan}{RGB}{0,166,235}
\definecolor{DESYorange}{RGB}{242,142,0}
\definecolor{DESYgray}{RGB}{119,119,119}
\definecolor{bjetpurple}{RGB}{117,112,179}

\usepackage[compat=1.1.0]{tikz-feynman}
\usepackage{contour}
\usetikzlibrary{arrows,shapes,positioning}
\usetikzlibrary{decorations.markings}
\usetikzlibrary{decorations.pathmorphing}
\usetikzlibrary{decorations.pathreplacing}
\usetikzlibrary{patterns}
\usetikzlibrary{plotmarks}
\usetikzlibrary{shadows}

\usepackage[margin=8mm,font=small,labelfont=bf,format=plain]{caption}
\usepackage[margin=8mm,font=small,labelfont=bf,format=plain]{subcaption}

\usepackage{multirow} %

\newcommand{\subfigref}[1]{({\protect\subref{#1}})}

\usepackage[
  bookmarks,                   %
  bookmarksopenlevel=1,        %
  bookmarksnumbered=true,      %
  pdfstartpage={1},            %
  pdfstartview={FitH},         %
  pdfkeywords={},              %
  pdfsubject={},               %
  pdfcreator={LaTeX with KOMA-Script and hyperref package},
  hyperfootnotes=true,         %
  linkbordercolor={0 1 1},     %
  menubordercolor={0 1 1},     %
  urlbordercolor={1 0 0}       %
]{hyperref}                    %

\usepackage{slashed}

\newcount\colveccount
\newcommand*\colvec[1]{
        \global\colveccount#1
        \begin{pmatrix}
        \colvecnext
}
\def\colvecnext#1{
        #1
        \global\advance\colveccount-1
        \ifnum\colveccount>0
                \\
                \expandafter\colvecnext
        \else
                \end{pmatrix}
        \fi
}

\usepackage{lmodern}
\usepackage{listings}
\usepackage[english]{babel}%
\usepackage{csquotes}%
\usepackage[style=numeric-comp,sorting=none]{biblatex}
\newcommand{\customprintbibliography}{%
  \printbibliography[heading=none]
  \newpage
}

\usepackage{amssymb}

\newcommand{\eP}{e^{+}}
\newcommand{\eM}{e^{-}}

\newcommand{\nubar}{\bar{\nu}}

\newcommand{\qbar}{\bar{q}}

\title{ILD benchmark: Quartic Gauge Couplings}
\author[1,2]{Jakob Beyer}
\author[1]{Jenny List}
\affil[1]{Deutsches Elektronen-Synchroton (DESY), Hamburg, Germany}
\affil[2]{Universit\"at Hamburg, Hamburg, Germany}

\begin{document}

\section{Introduction}
The quartic gauge coupling (QGC) of four boson of the electro-weak interaction relies on the Standard Model Higgs in order to not violate unitarity at high energies.
Beyond the SM (BSM) physics which affects the Higgs is therefore well-motivated to appear in the scattering of such bosons called Vector Boson Scattering (VBS).
A high center-of-mass energy at the TeV scale will be necessary to observe such deviations.

High-energy hadron colliders such as the LHC can measure VBS but are limited by high rates of pile-up background and secondary interactions.
This makes measurements in hadronic final states challenging and precision measurements must rely on leptonic and semi-leptonic final states. 
In the low-background environment of an electron-positron collider fully hadronic final states can be used for precision measurements.

Here, the measurement of the $\nu\nubar+q\qbar q\qbar$ final state is studied in the context of a QGC analysis at ILD for a $\sqrt{s}=1\,$TeV ILC.
Particular focus is set on identification of critical reconstruction steps and possible improvements.

\section{Simulation \& analysis setup}
\begin{figure} 
  \centering
  \includegraphics[scale=1.0]{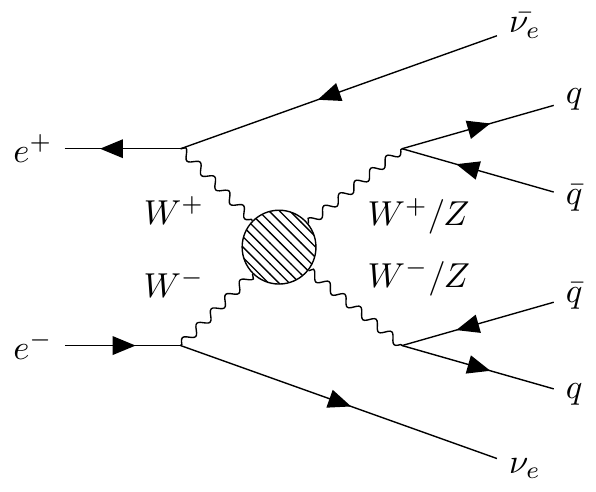}
  \caption{Diagram of vector boson scattering in the $\nu\nubar q\qbar q\qbar$ final state at $\eP\eM$ colliders.}
  \label{FEY:SignalProcess}
\end{figure}

An analysis of the signal event reconstruction requires a definition of signal events.
Samples produced for ILD are not created using specifically on-shell bosons and do not contain such massive bosons in their generator level history.
The definition of a signal event therefore uses the helicity of the initial particles, the flavours of the final state particles and invariant masses of the quark-antiquark and neutrino-antineutrino pairs to select a signal like topology (see fig.\ \ref{FEY:SignalProcess}).

\begin{itemize}
  \item \textbf{Incoming particles:} 
    Since both initial particles should radiate a $W$ boson the incoming $\eM$ must be left-handed and the incoming $\eP$ must be right-handed.
  \item \textbf{Final state flavours:} 
    Because both neutrinos should originate from a $e \rightarrow W\nu$ vertex they must both be electron-like.
    The four quark final state is supposed to correspond to a $WW$ or a $ZZ$ intermediate state. 
    All possible $(q\qbar)(q\qbar)$ double-pairs are considered and categorized as possible $ZZ$ candidate if both quarks in each pair are of same flavour or as possible $WW$ candidate if both pairs consist of a up-type down-type combination.
  \item \textbf{Boson candidate invariant masses:} 
    Mass windows are imposed on the possible $VV$ ($V=W/Z$) candidates to only include double-resonant events in the signal definition. 
    A $(q_1\qbar_2)(q_3\qbar_4)$ double-pair must fulfill the following conditions to be considered a proper $ZZ$ ($WW$) candidate. 
    The sum of both invariant masses has to fall into the range of $171 (147)\,\text{GeV} < m_{inv}^{1,2} + m_{inv}^{3,4} < 195 (171)\,\text{GeV}$ and their difference must not exceed $20\,\text{GeV} > | m_{inv}^{1,2} - m_{inv}^{3,4} |$.
    If more than one proper candidate double-pair is found the double-pair with the smallest mass difference is chosen.
  \item \textbf{Neutrino pair invariant mass:} 
    To exclude events originating from a triple-resonant $WWZ$ final state from the signal definition the neutrino pair must have an invariant mass of more than $100\,\text{GeV} > m_{\nu\nubar}$.\footnote{$WWZ$ can on tree level originate from a quartic gauge boson interaction $Z\rightarrow ZWW$ but this process is suppressed by the far off-shell nature of the first $Z$.}
\end{itemize}

For the signal event data set of the present analysis the \texttt{WHIZARD1.95} generator files of the DBD studies are run through the current ILD detector simulation and event reconstruction in the central ILD production.
This production is based on \texttt{iLCSoft v02-00-02} and \texttt{ILDConfig v02-00-02}.
Estimates of the beam-induced backgrounds are updated and include - for the first time - $e^+e^-$ pair background.
A large (\texttt{ILD\_l5\_o1\_v02}/\texttt{IDR-L}) and a small  (\texttt{ILD\_s5\_o1\_v02}/\texttt{IDR-S}) detector model\footnote{Major differences between the models are the thickness of the TPC tracker and the magnetic field strength. The large model is largely similar to the model used in the TDR~\cite{Behnke:2013lya}. In the smaller model the TPC radial thickness is reduced and the magnetic field strength is increased.} are simulated in parallel and both reconstruction results are available for analysis.

The events are weighted to $1\,$ab$^{-1}$ of $1\,$TeV ILC running with polarizations of $P_{\eM} = -80\%$ and $P_{\eP} = +30\%$.

After the event reconstruction a simple analysis is performed to reconstruct two di-jet vector boson candidates. 
It consists of three steps.
\begin{itemize}
  \item \textbf{Beam background removal:}
    Beam-induced backgrounds are removed by applying an exclusive $k_t$ jet clustering algorithm to all reconstructed particles. 
    All particles which do not end up in one of the clustered jets are considered beam background and removed from further analysis. 
    Here, a cone parameter of 1.3 is chosen and the clustering is performed to four jets.
  \item \textbf{Jet clustering:}
    All remaining particles are clustered to four jets using exclusive $ee-k_t$ jet clustering.
  \item \textbf{Jet pairing:}
    The four jets are paired to two di-jet vector boson candidates by choosing the combination that yields the smallest difference between the vector boson masses $|m_{jj,1} - m_{jj,2}|$.
\end{itemize}
For the jet clustering steps the \texttt{Marlin} implementation of \texttt{FastJet} is used \cite{Cacciari:2011ma}.

Some of the results of the study presented here require the possibility of performing an idealized version of one of the reconstruction steps.
This is possible within the data set of the ILD production because the simulation and reconstruction are implemented to keep track of the connection between generator level particles and reconstructed particles.
With these connections idealized reconstruction steps can be performed by first finding the object of interest on generator level and then following the connections to the corresponding reconstructed particles.
The \texttt{TrueJet} tool \cite{MikaelBerggen2018} implements this functionality with a focus on jet reconstruction analyses and is employed in this study.

\section{Hadronic diboson reconstruction\normalsize\protect\footnote{The source code used for the analysis in this section can be found in~\cite{GitHubQGCILDbenchRepo}.}}
The goal of the search for VBS is to search for possible anomalous couplings.
In a general Effective Field Theory (EFT) framework these couplings generally differ between the $WW\rightarrow WW$ and the $WW \rightarrow ZZ$ vertices \cite{Fleper:2016frz}.
To set restrictions in the two (or more) dimensions of the anomalous coupling plane these vertices need to be disentangled in the measurement.
If the $WW$ and the $ZZ$ final states can be separated by the invariant masses of the decay products a separate measurement of the two vertices becomes trivial.
Therefore, the kinematic separation of the two final state is the main aspect of the analysis and the focus of this section.

\begin{figure}
  \centering
  \begin{subfigure}[t]{0.5\textwidth}
    \centering
    \includegraphics[width=\textwidth]{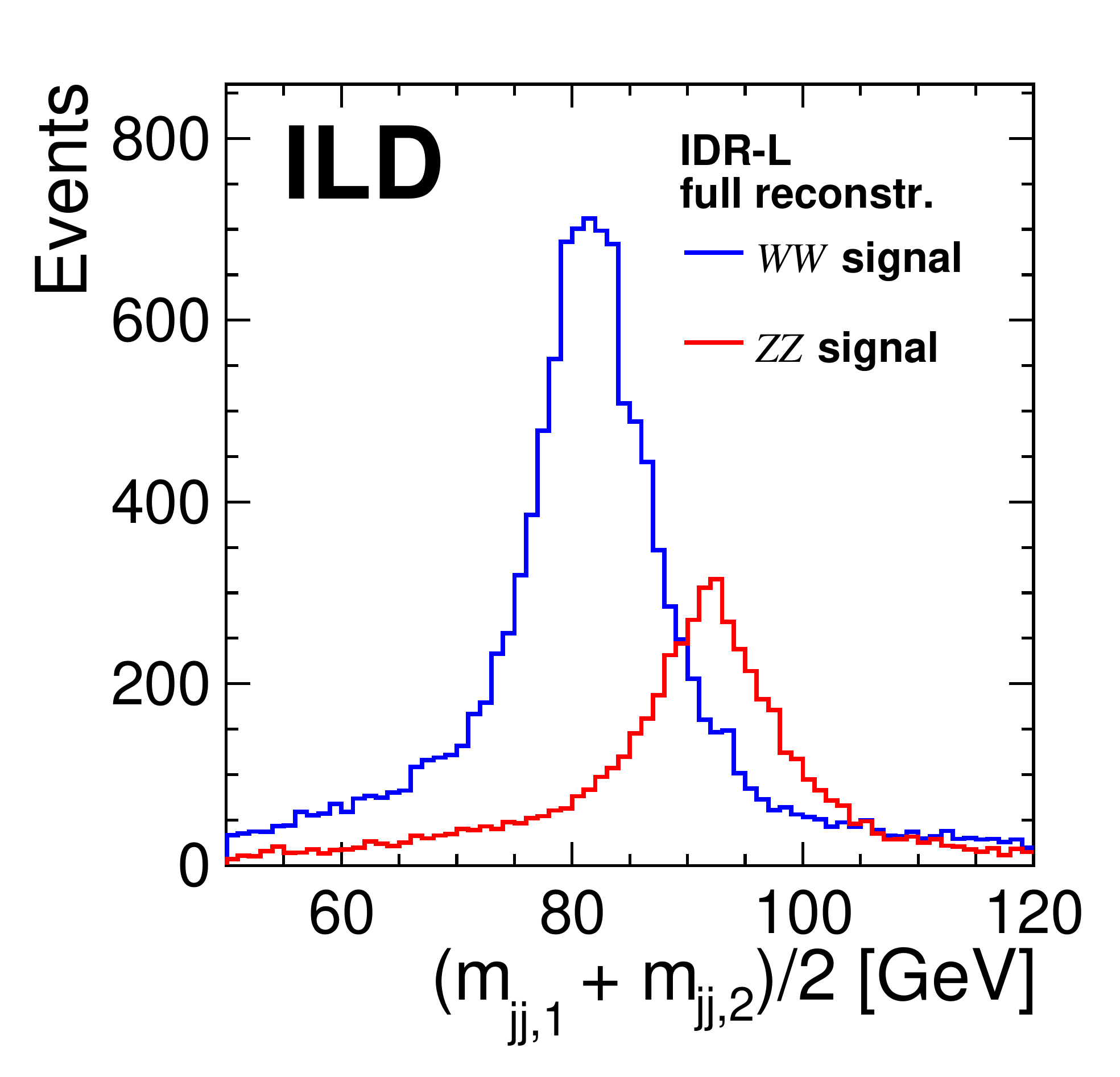}
    \caption{}
    \label{SUBFIG:IDRL_m}
  \end{subfigure}%
  \begin{subfigure}[t]{0.5\textwidth}
    \centering
    \includegraphics[width=\textwidth]{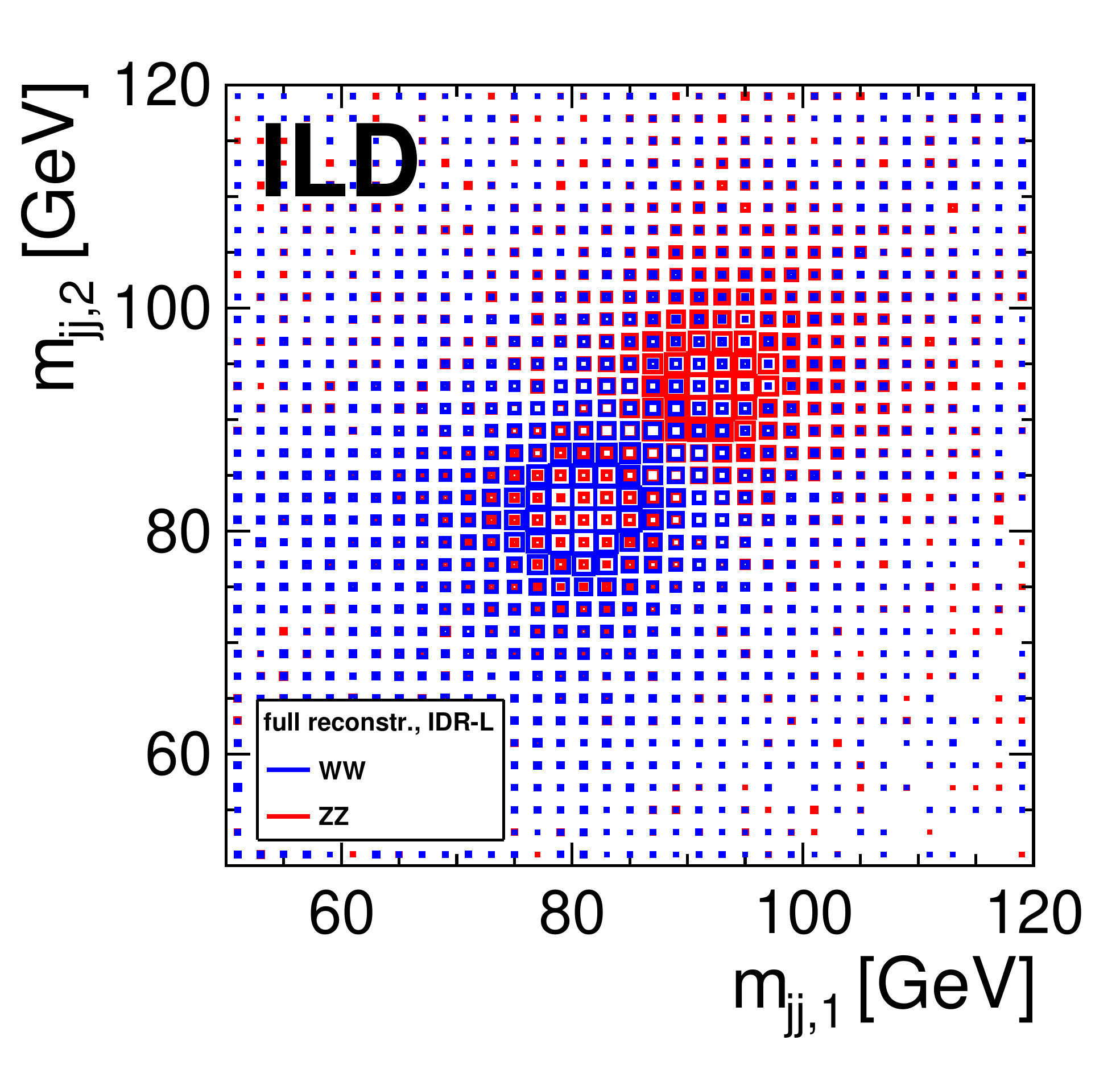}
    \caption{}
    \label{SUBFIG:IDRL_m_m}
  \end{subfigure}
  \caption{
    Reconstructed vector boson candidate mass spectra from full simulation in the large ILD model.
    \subfigref{SUBFIG:IDRL_m} Average of both masses, weighted to nominal luminosity of $1\,$ab$^{-1}$ of $1\,$TeV ILC running. 
    \subfigref{SUBFIG:IDRL_m_m} Mass spectrum for both vector boson candidates, with $WW$ and $ZZ$ spectra individually normalized.
  }
  \label{FIG:MassPlotsIDRL}
\end{figure}

Distributions of the reconstructed masses of the vector boson candidates are used to quantify this separation.
A strong overlap of the reconstructed $WW$ and $ZZ$ mass distributions with tails far into the high and low mass regions is observed when using the full analysis described above (fig.~\ref{FIG:MassPlotsIDRL}).
The question arises which effects are responsible for this overlap and how they might be mitigated.
By isolating one of the reconstruction aspects and performing it idealized it is possible to observe its significance.
Here, the \texttt{TrueJet} tool is used to create the mass distributions at different levels of idealized reconstruction.
This idealization is performed additive, i.e. each of the following levels includes the idealized steps of all the previous.

\begin{itemize}
  \item \textbf{Full reconstruction:} 
    All signal events and reconstruction steps as described in the previous section are used.
  \item \textbf{Cheated overlay:}\footnote{\textit{Overlay} is a term used to describe beam-induced backgrounds at ILC and originates from the way the background is overlayed on the event in the simulation.} 
    Instead of using exclusive jet clustering for beam-background removal it is performed by identifying the generator level background particles and removing the according reconstructed particles.
  \item \textbf{Cheated jets:} 
    The $ee-k_t$ jet clustering is replaced an ideal clustering which associates the reconstructed particles with the particles of the generator level jet.
  \item \textbf{Cheated bosons:} 
    While no on-shell bosons are included in the generator level it is possible to find di-quark candidates using the signal definition described in the previous section.
    These di-quarks can then be associated with the generator level jets, thereby performing an idealized pairing of into di-jet vector boson candidates. 
  \item \textbf{No semi-leptonic events:} 
    Semi-leptonic decays of heavy hadrons in jets produce additional missing four-momentum in the jet and worsens the jet energy resolution. 
    Events with such decays in jets are identified and excluded for this level.
\end{itemize}

\begin{figure}
  \centering
  \begin{subfigure}[t]{0.5\textwidth}
    \centering
    \includegraphics[width=\textwidth]{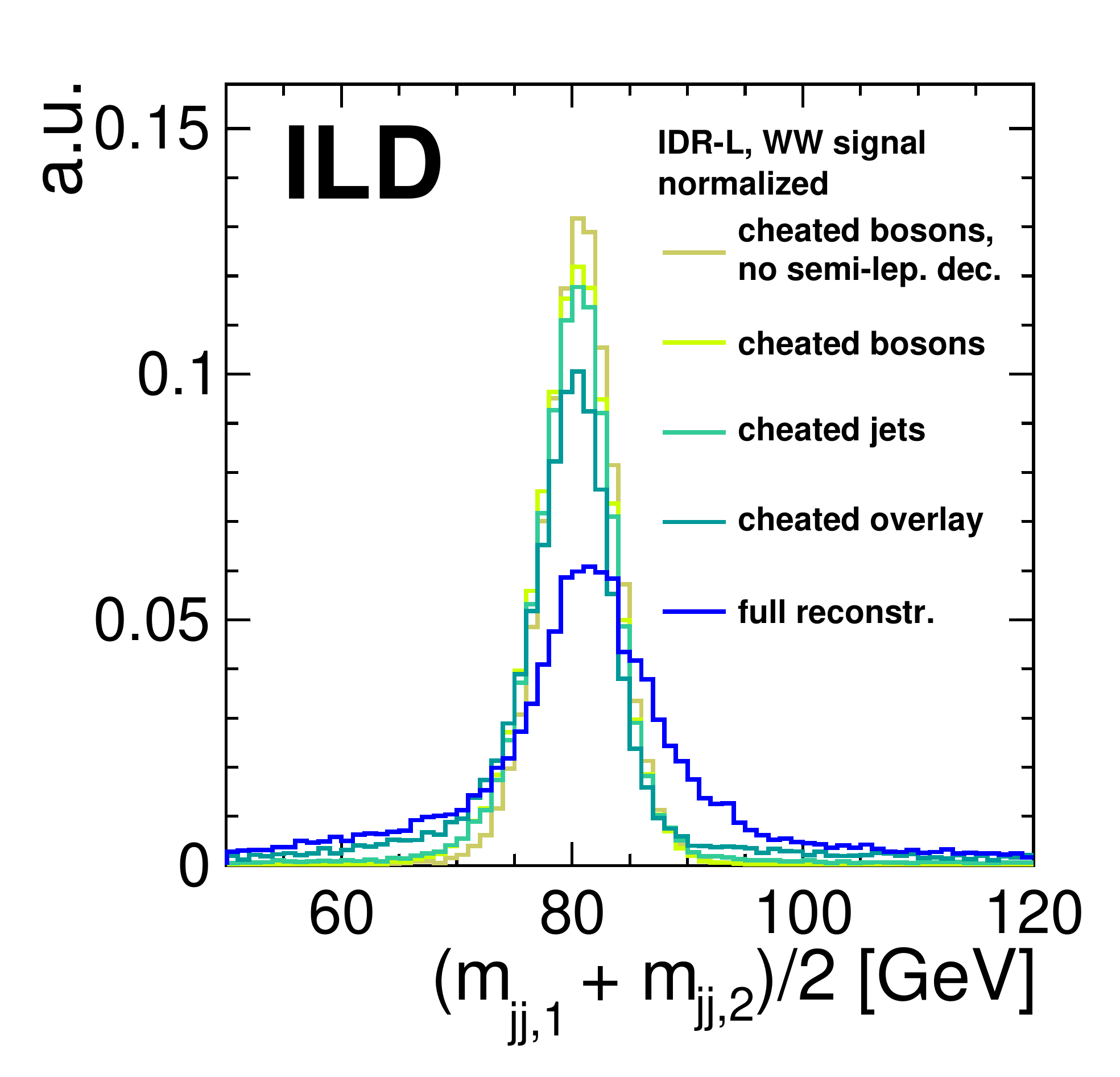}
    \caption{}
    \label{SUBFIG:LevelComparison_WW}
  \end{subfigure}%
  \begin{subfigure}[t]{0.5\textwidth}
    \centering
    \includegraphics[width=\textwidth]{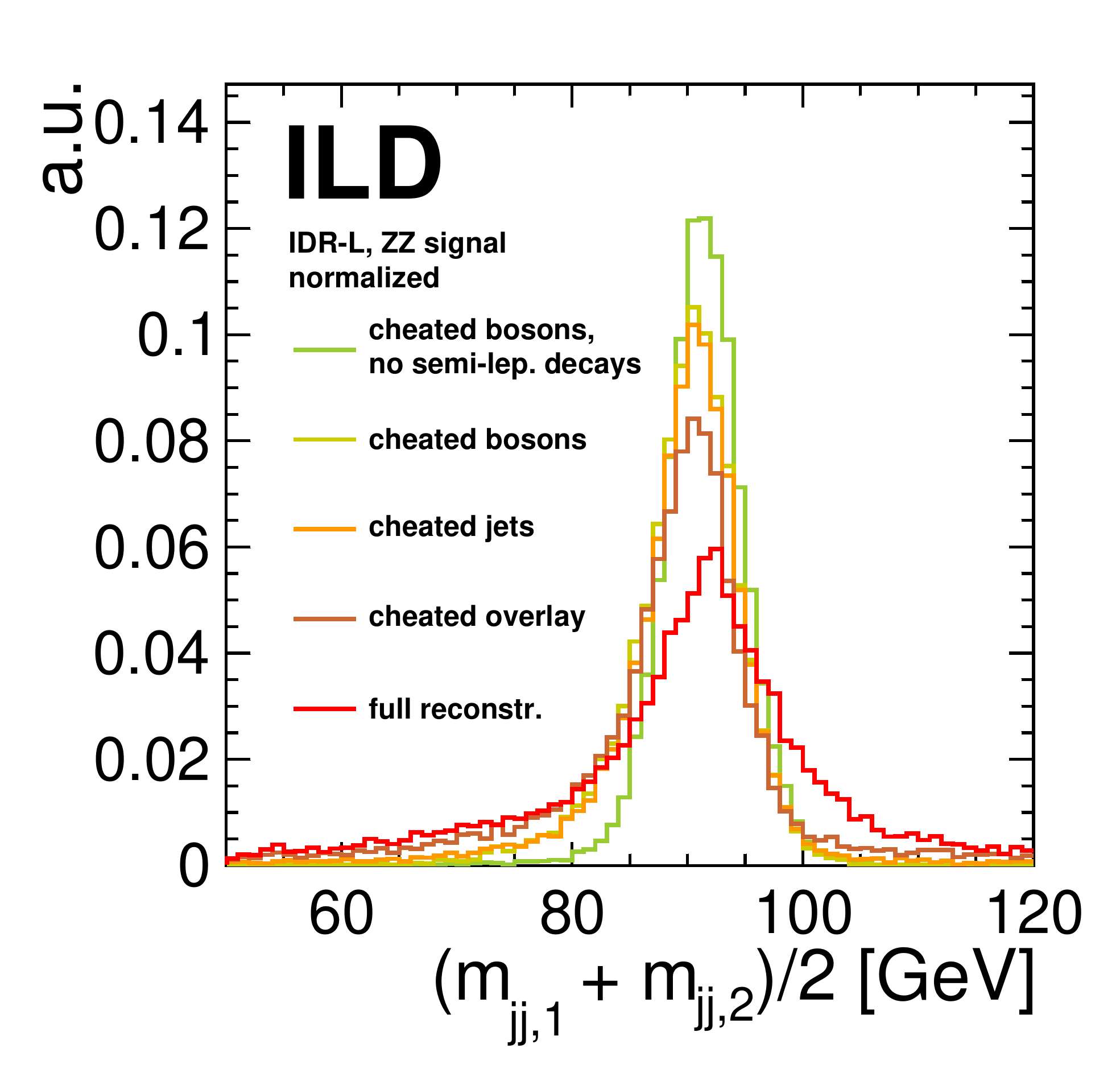}
    \caption{}
    \label{SUBFIG:LevelComparison_ZZ}
  \end{subfigure}
  \caption{
    Normalized \subfigref{SUBFIG:LevelComparison_WW} $WW$ and \subfigref{SUBFIG:LevelComparison_ZZ} $ZZ$ mass average distributions for the two reconstructed vector boson candidates at different levels of idealized reconstruction in the large ILD model.
    Each level is individually normalized to 1.
  }
  \label{FIG:LevelComparison}
\end{figure}

\begin{figure}
  \centering
  \begin{subfigure}[t]{0.5\textwidth}
    \centering
    \includegraphics[width=\textwidth]{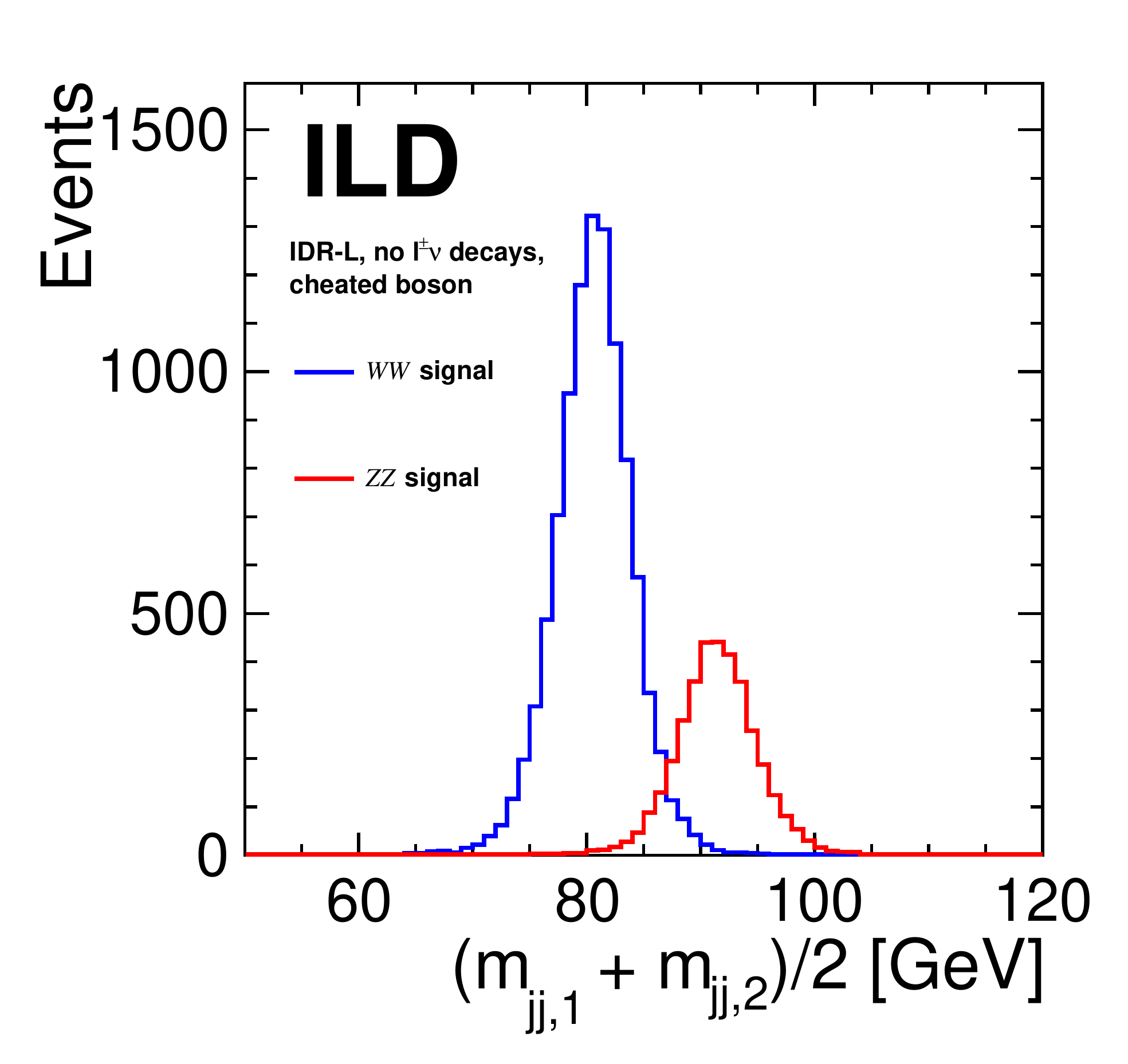}
    \caption{}
    \label{SUBFIG:IDRL_m_icn_noSLD}
  \end{subfigure}%
  \begin{subfigure}[t]{0.5\textwidth}
    \centering
    \includegraphics[width=\textwidth]{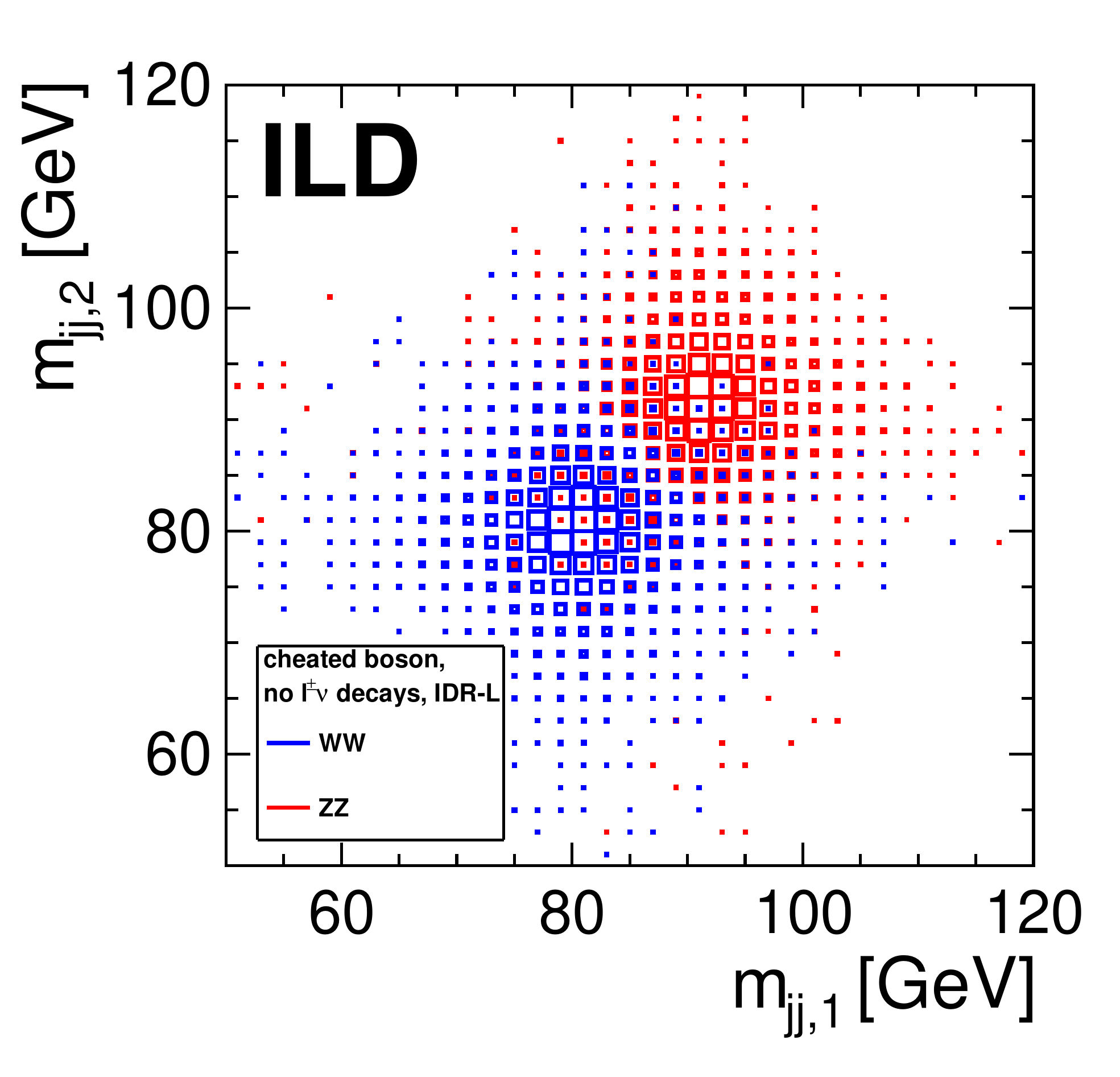}
    \caption{}
    \label{SUBFIG:IDRL_m_m_icn_noSLD}
  \end{subfigure}
  \caption{
    Reconstructed vector boson candidate mass spectra from full simulation in the large ILD model with idealized high level reconstruction and only for events without semi-leptonic decays in jets.
    \subfigref{SUBFIG:IDRL_m_icn_noSLD} Average of both masses, weighted to nominal luminosity of $1\,$ab$^{-1}$ of $1\,$TeV ILC running. 
    \subfigref{SUBFIG:IDRL_m_m_icn_noSLD} Mass spectrum for both vector boson candidates, with $WW$ and $ZZ$ spectra individually normalized.
  }
  \label{FIG:MassPlotsIDRLcheated}
\end{figure}

A comparison of the mass distributions at these different levels shows that the most limiting reconstruction aspects are the removal of beam-backgrounds, the jet clustering and the treatment of semi-leptonic decays within jets (fig.~\ref{FIG:LevelComparison}).
When the treatment of all of these aspects is idealized only a small overlap between the mass distributions remains (fig.~\ref{FIG:MassPlotsIDRLcheated}).
These aspects deserve dedicated studies aiming to improve each associated algorithm.
Here, only the correction of semi-leptonic decays will be discussed in some detail in the section below.

\begin{figure}
  \centering
  \begin{subfigure}[t]{0.5\textwidth}
    \centering
    \includegraphics[width=\textwidth]{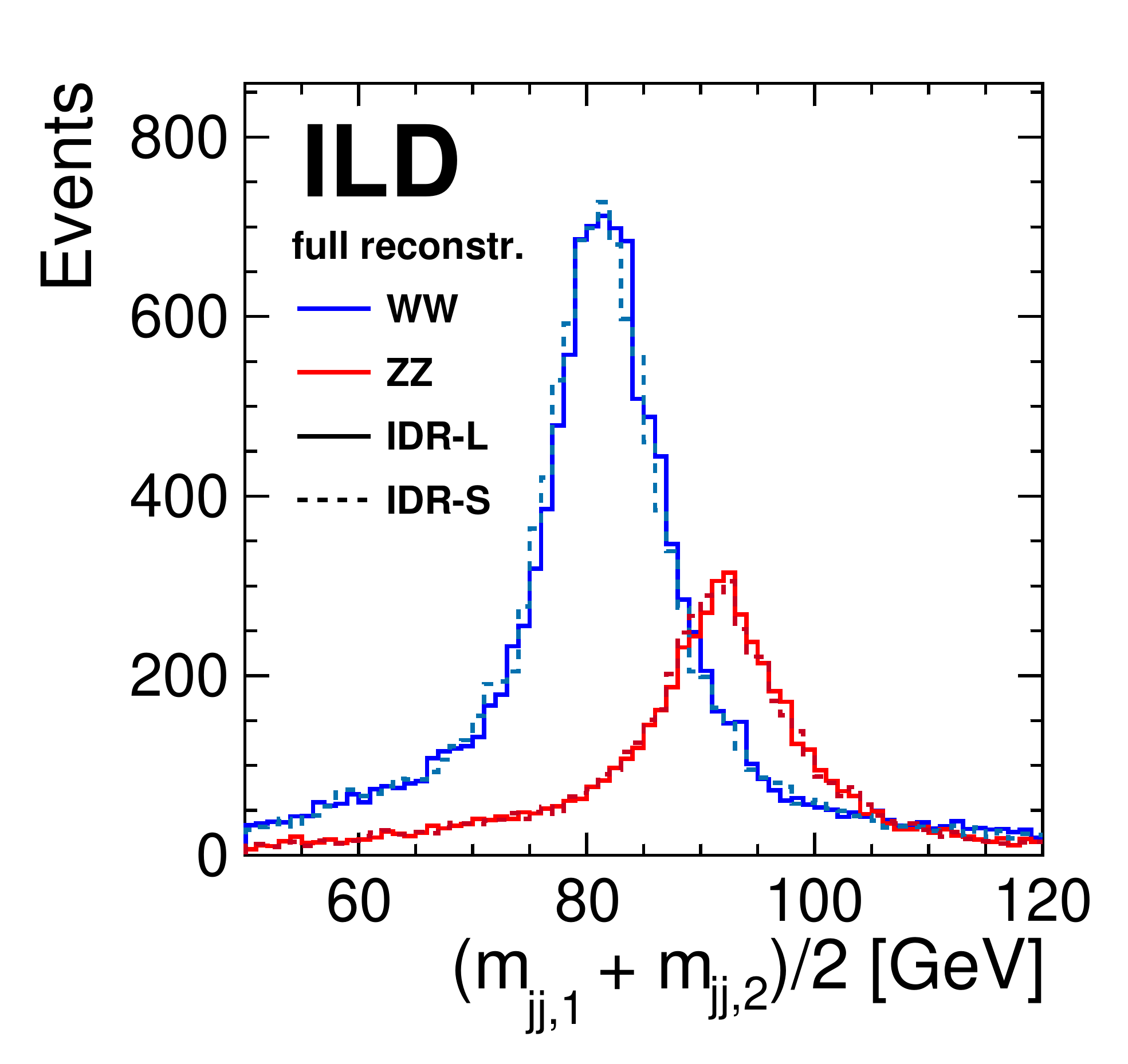}
    \caption{}
    \label{SUBFIG:DetectorComparison_rec}
  \end{subfigure}%
  \begin{subfigure}[t]{0.5\textwidth}
    \centering
    \includegraphics[width=\textwidth]{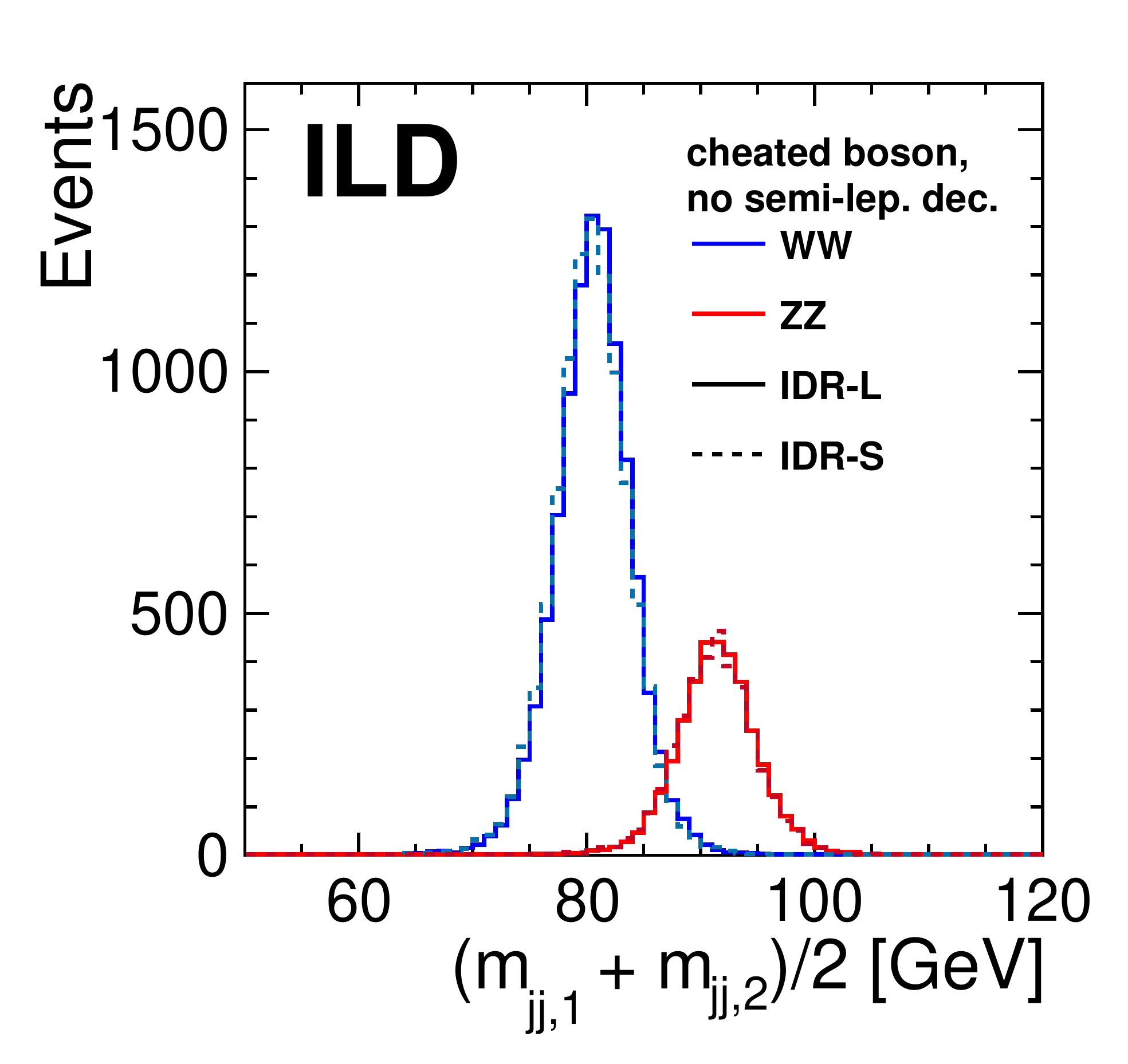}
    \caption{}
    \label{SUBFIG:DetectorComparison_icn}
  \end{subfigure}%
  \caption{    
    Reconstructed average mass distributions of the reconstructed vector boson candidates from full simulation in the large and small ILD models.
    Events are weighted to nominal luminosity of $1\,$ab$^{-1}$ of $1\,$TeV ILC running. 
    \subfigref{SUBFIG:DetectorComparison_rec} Using the full high level reconstruction and all signal events.
    \subfigref{SUBFIG:DetectorComparison_icn} With idealized high level reconstruction and only for signal events without semi-leptonic decays in jets.
  }
  \label{FIG:DetectorComparison}
\end{figure}

To aid in the optimization of ILD the separation performance is investigated for the large and small ILD models.
When using the full reconstruction chain described in the previous section no significant difference is observed between the models (fig.~\ref{SUBFIG:DetectorComparison_icn}).
Because smaller difference in the detector performance may be hidden by high level reconstruction effects the comparison is repeated using most idealized high level reconstruction (fig.~\ref{SUBFIG:DetectorComparison_icn}).
At this level the only remaining effects after fragmentation and hadronization are those of the hardware response and the Particle Flow algorithm.
Both detectors again show very similar distributions.
This can be explained by low average single jet energy.
For jet energies below $100\,$GeV the difference of the jet energy resolution (JER) between the detector models is small~\cite{LCWS18JER}.

\subsection{Reconstruction of high-$\boldsymbol{m_{VV}}$ events\normalsize\protect\footnote{Additional distributions are given in the appendix.}}

  In an EFT-framework new physics is assumed to be at high energies and out of direct reach of the collider experiment.
  Sensitivity to the EFT couplings rises with the center-of-mass energy in the affected scattering process as it comes closer to the assumed BSM physics.
  VBS events with a high di-boson invariant mass $m_{VV}$ are therefore of special relevance.
  
\begin{figure}
    \centering
    \begin{subfigure}[t]{0.5\textwidth}
      \centering
      \includegraphics[width=\textwidth]{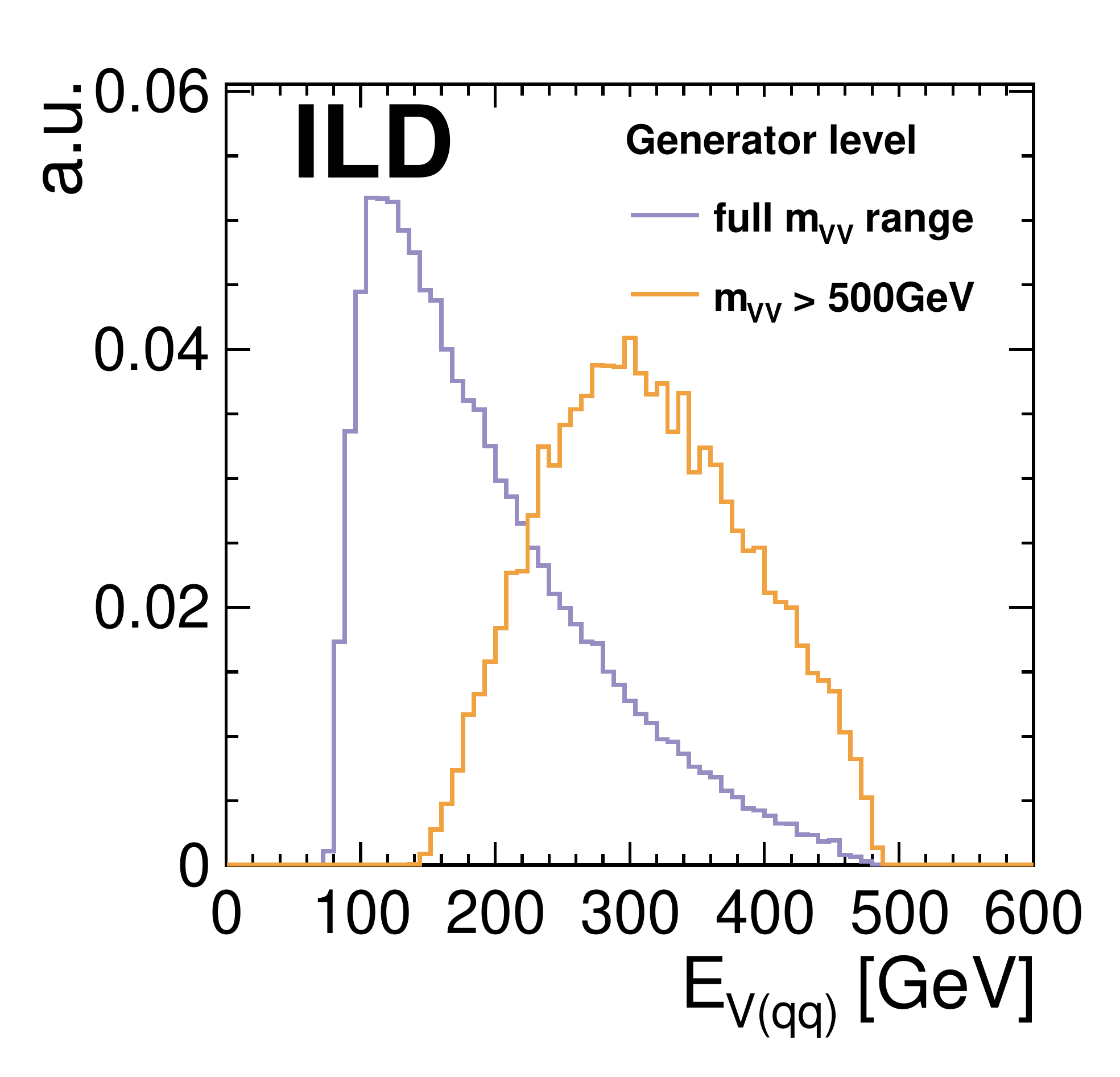}
      \caption{}
      \label{SUBFIG:SampleComparison_E}
    \end{subfigure}%
    \begin{subfigure}[t]{0.5\textwidth}
      \centering
      \includegraphics[width=\textwidth]{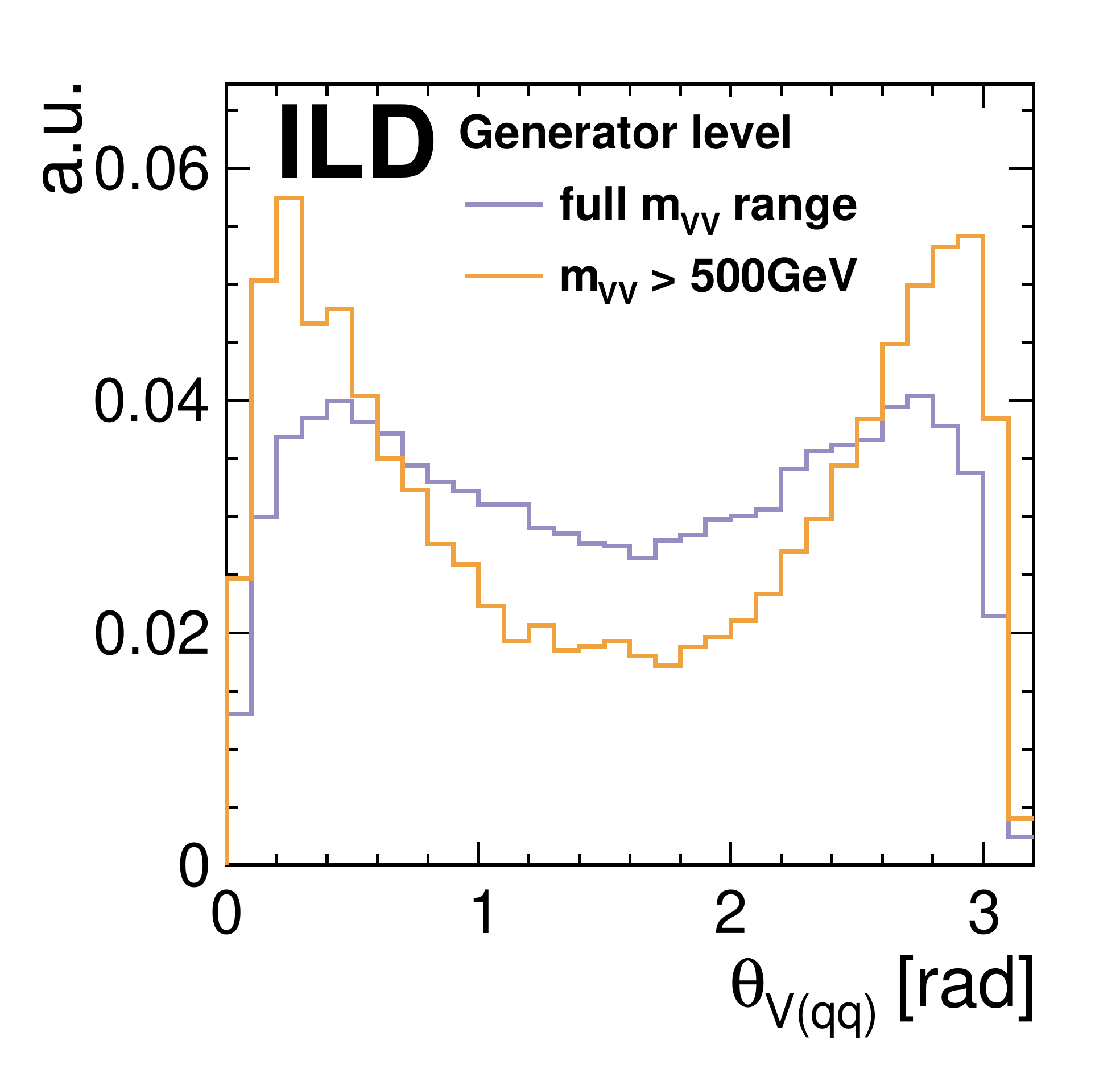}
      \caption{}
      \label{SUBFIG:SampleComparison_theta}
    \end{subfigure}%
    \caption{    
      Normalized kinematic distributions of the generator level di-quark bosons for the full-$m_{VV}$ range dataset and the dataset with an $m_{VV}>500\,$GeV restriction on generator level.
      Two bosons are filled per event.
      \subfigref{SUBFIG:SampleComparison_E} Energy of the di-quark boson.
      \subfigref{SUBFIG:SampleComparison_theta} Polar angle of the di-quark bosons.
    }
    \label{FIG:SampleComparison}
  \end{figure}

  An additional dataset with a generator level cut of $m_{VV}>500\,$GeV is produced to investigate the reconstruction in this kinematic region.
  Due to the phase space restriction the signal cross section for $WW$ ($ZZ$) events reduces from $21.9\,$fb ($10.1\,$fb) to $1.37\,$fb ($0.53\,$fb).
  The produced bosons have a stronger boost and are tend to fly into more forward regions (see fig.\ \ref{FIG:SampleComparison}).
  Beam-background and detector effects are dependent on angle and energy of the particle.
  Accordingly, a change in dominant reconstruction effects can be expected, potentially accompanied by differences between the large and small detector model.

\begin{figure}
    \centering
    \begin{subfigure}[t]{0.5\textwidth}
      \centering
      \includegraphics[width=\textwidth]{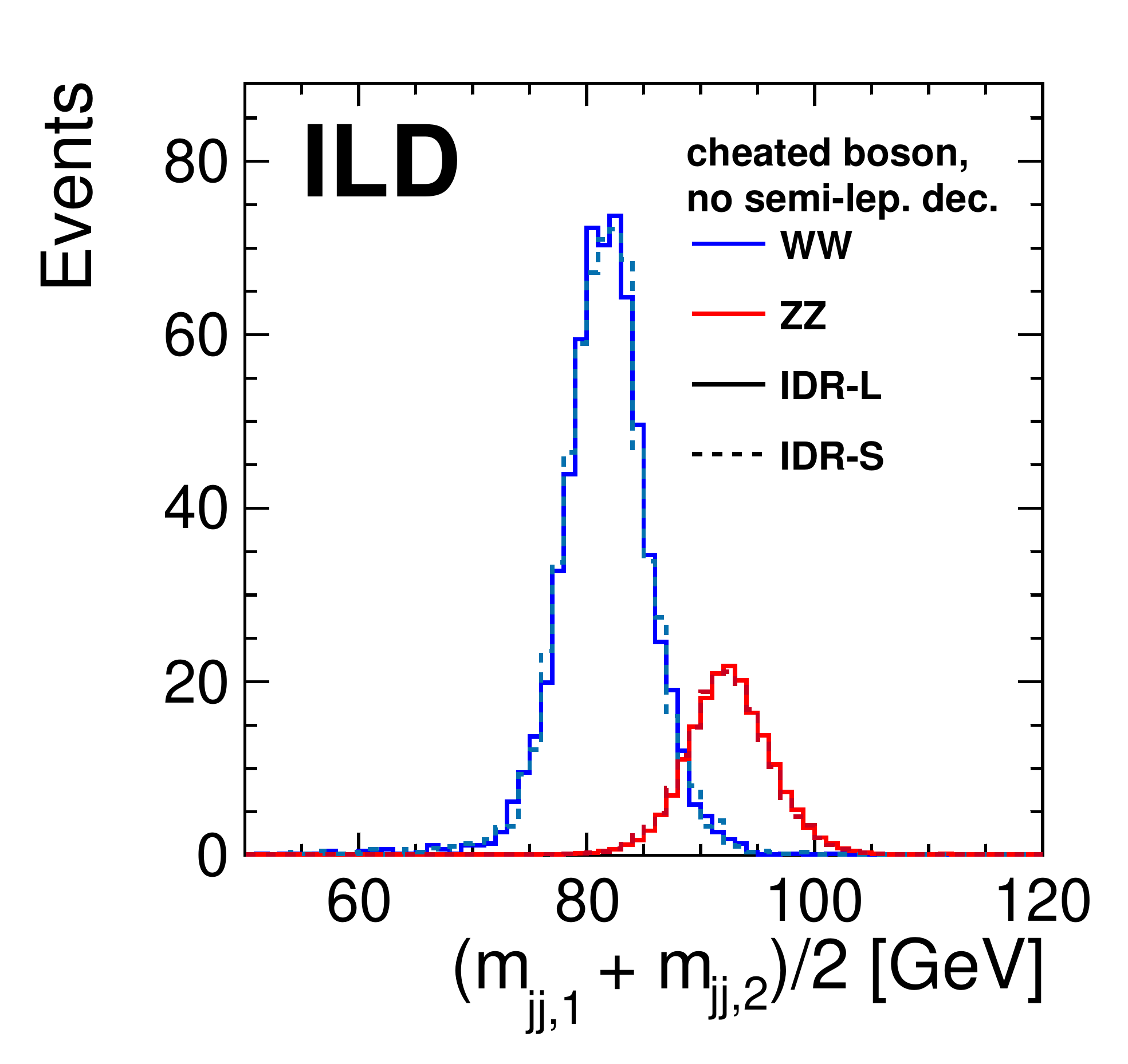}
      \caption{}
      \label{SUBFIG:DetectorComparison_icn_highQ2}
    \end{subfigure}%
    \begin{subfigure}[t]{0.5\textwidth}
      \centering
      \includegraphics[width=\textwidth]{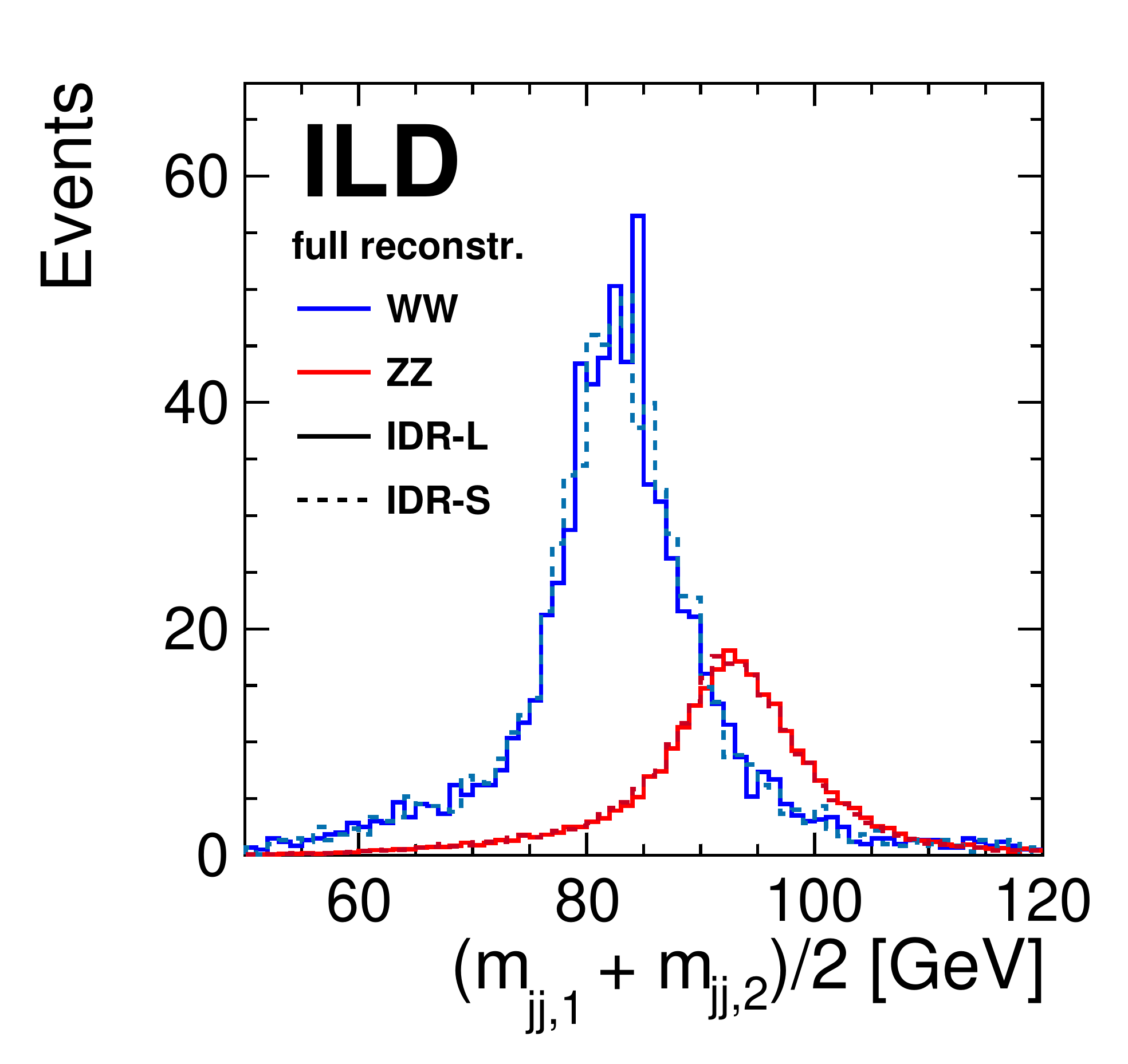}
      \caption{}
      \label{SUBFIG:DetectorComparison_rec_highQ2}
    \end{subfigure}%
    \caption{    
      Average mass distributions of the reconstructed vector boson candidates from full simulation in the large and small ILD models with the dataset with a $m_{VV}>500\,$GeV restriction on generator level.
      Events are weighted to nominal luminosity of $1\,$ab$^{-1}$ of $1\,$TeV ILC running. 
      \subfigref{SUBFIG:DetectorComparison_icn_highQ2} With idealized high level reconstruction and only for signal events without semi-leptonic decays in jets.
      \subfigref{SUBFIG:DetectorComparison_rec_highQ2} Using the full high level reconstruction and all signal events.
    }
    \label{FIG:DetectorComparison_highQ2}
  \end{figure}

  When cheating all high-level reconstruction aspects the detector models both show comparable performance as for the full $m_{VV}$ range (see fig.\ \ref{SUBFIG:DetectorComparison_icn_highQ2}).
  At this level the primary detector effect is the JER, whose energy and angular dependence are known~\cite{LCWS18JER}.
  On average a higher jet energy leads to improved JER and a better relative performance of the large model.
  However, for jets going further into the forward direction the JER decreases and leads to a better relative JER for the small model due to the higher $B$-field.
  From the full high-$m_{VV}$ range to the high-$m_{VV}$ signal events the kinematic distributions changed such that energy- and angular-dependent changes approximately cancel. 
  In addition, the high-$m_{VV}$ signal events primarily hit a detector region with comparable JER-performances for both models.

\begin{figure}
    \centering
    \begin{subfigure}[t]{0.5\textwidth}
      \centering
      \includegraphics[width=\textwidth]{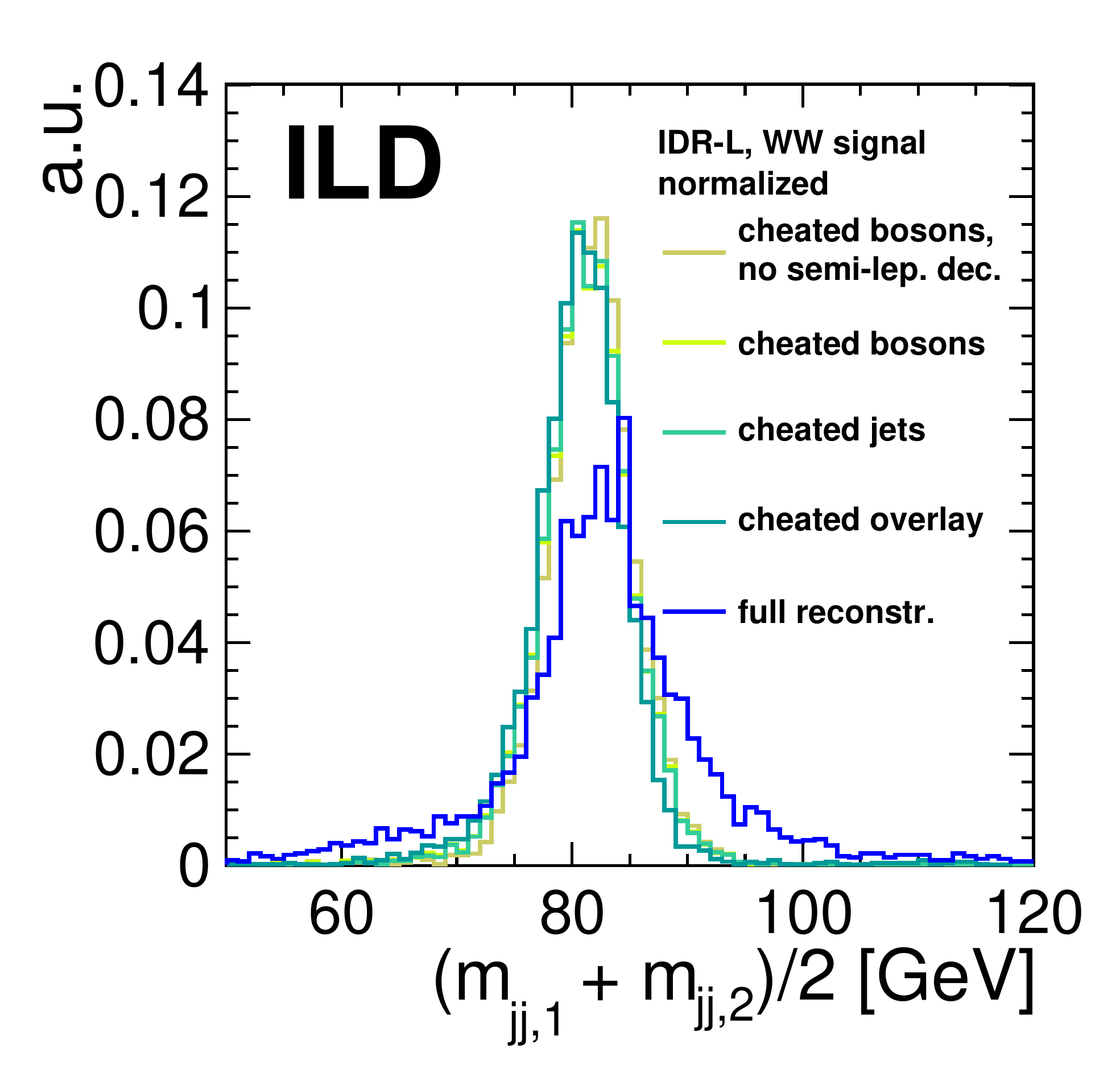}
      \caption{}
      \label{SUBFIG:LevelComparison_WW_highQ2}
    \end{subfigure}%
    \begin{subfigure}[t]{0.5\textwidth}
      \centering
      \includegraphics[width=\textwidth]{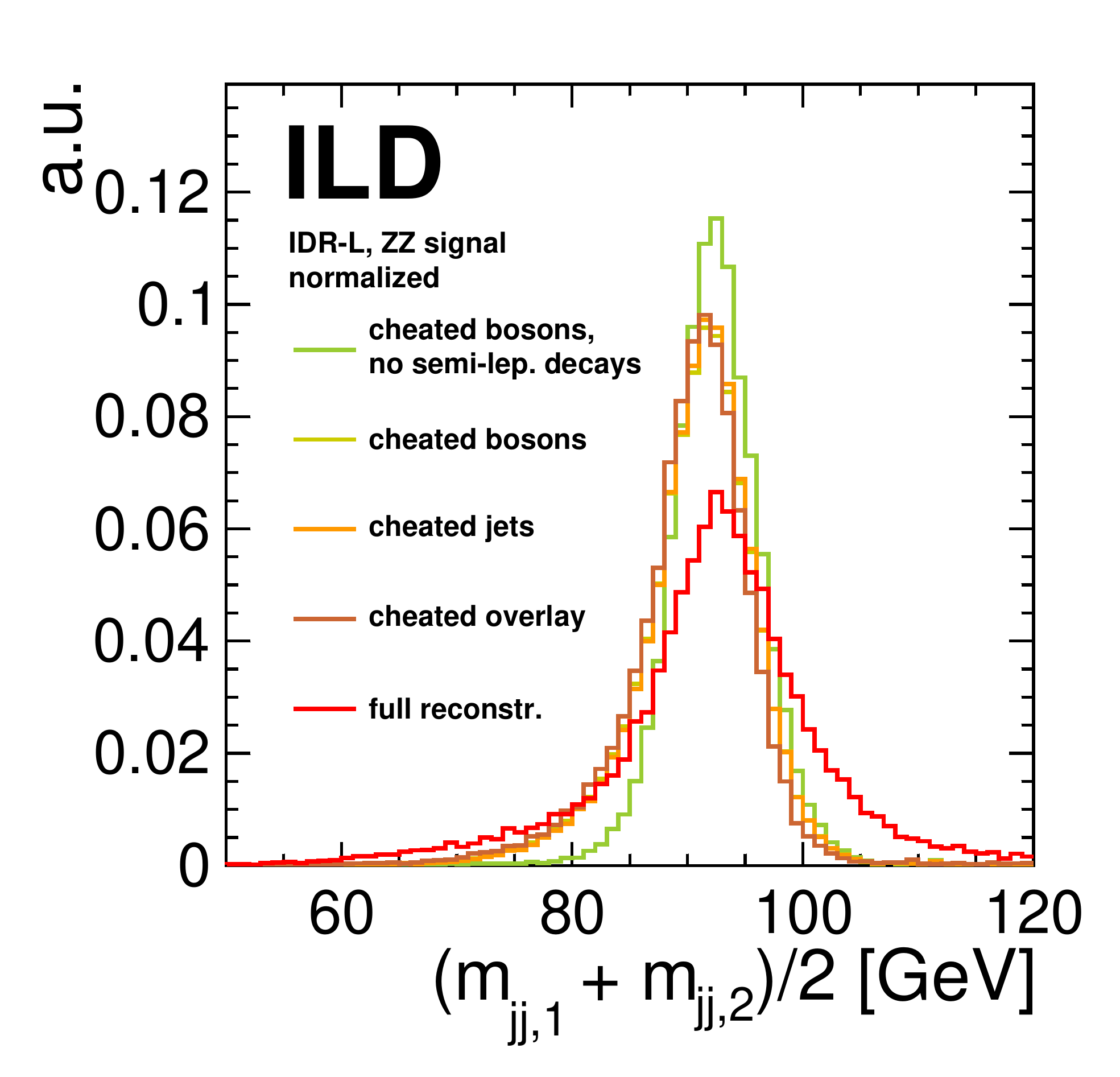}
      \caption{}
      \label{SUBFIG:LevelComparison_ZZ_highQ2}
    \end{subfigure}%
    \caption{    
      Normalized \subfigref{SUBFIG:LevelComparison_WW_highQ2} $WW$ and \subfigref{SUBFIG:LevelComparison_ZZ_highQ2} $ZZ$ mass average distributions for the two reconstructed vector boson candidates at different levels of idealized reconstruction in the large ILD model.
      The dataset with a $m_{VV}>500\,$GeV restriction on generator level is used.
      Each level is individually normalized to 1.
    }
    \label{FIG:LevelComparison_highQ2}
  \end{figure}

  After the full high-level reconstruction both models still show results very similar to the full $m_{VV}$ range dataset (see fig.\ \ref{SUBFIG:DetectorComparison_rec_highQ2}).
  This is despite an observed shift in reconstruction effects (fig.\ \ref{FIG:LevelComparison_highQ2}).
  Jet clustering and di-jet pairing to bosons only have a small influence on the mass distributions.
  Due to strongly boosted bosons and the large opening angle between them it is not difficult to cluster particles into two boson candidates.
  In their place the removal of beam-backgrounds takes on a significantly increased role.
  The decay products of the forward-produced bosons are more likely to overlap with the forward-focused beam-background particles in the calorimeters.
  While the easier clustering environment would allow for a better boson mass resolution this effect is approximately compensated by the increased beam-background contribution.
  
  In the specific context of an EFT-analysis of the VBS process the limiting factors are found to be beam-backgrounds and semi-leptonic decays.
  Jet clustering in this kinematic region is no longer a dominating factor.

\subsection{Quantifying $\boldsymbol{WW/ZZ}$ separation}

  So far the kinematic $WW/ZZ$ separation was addressed using the reconstructed mass distributions.
  A quantitative method to describe this separation may help to assess its obstacles and guide possible improvements in future studies.
  Here, the percentage of correctly identified $VV$ pairs ($V=W/Z$) is proposed as such a figure of merit and constructed in following way.
  
  A simple separation cut $m_{cut}$ is introduced in the one dimensional projection $\left( m_{jj,1} + m_{jj,2}\right) / 2$ of the two reconstructed boson masses (e.g. fig.~\ref{SUBFIG:IDRL_m}).
  All events falling below $m_{cut}$ are classified as reconstructed $WW$ and all above $m_{cut}$ as reconstructed $ZZ$.
  The cut value $m_{cut}$ is then scanned over the reconstructed mass range.
  At each $m_{cut}$ point the efficiencies $\epsilon_{WW} \left(\epsilon_{ZZ}\right)$ to correctly identify a $WW (ZZ)$ pair are extracted.
  The set of all $\left(\epsilon_{WW},\epsilon_{ZZ}\right)$ pair points forms a receiver operating characteristic (ROC) curve (fig.~\ref{FIG:IDRSepCurves}\footnote{Figures for the high-$m_{VV}$ dataset are shown in the appendix.}).
  Here, the characteristic $WW/ZZ$ separation efficiency is defined as the point of the ROC curve where the efficiencies for correct $WW$ and $ZZ$ identification are equal ($\epsilon_{WW}=\epsilon_{ZZ}$).

  \begin{figure}
    \centering
    \begin{subfigure}[t]{0.5\textwidth}
      \centering
      \includegraphics[width=\textwidth]{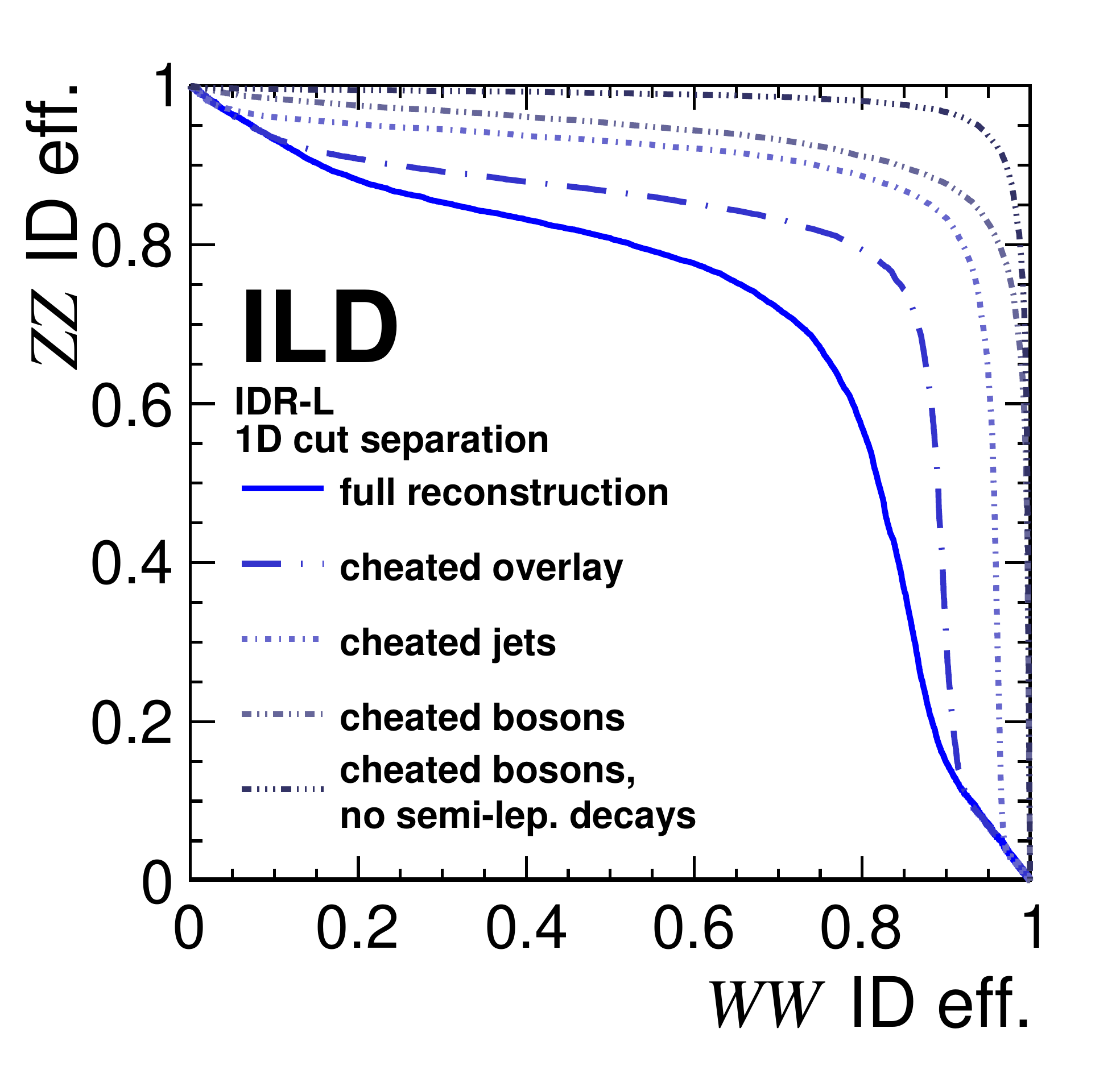}
      \caption{}
      \label{SUBFIG:IDRL_sep_curves}
    \end{subfigure}%
    \begin{subfigure}[t]{0.5\textwidth}
      \centering
      \includegraphics[width=\textwidth]{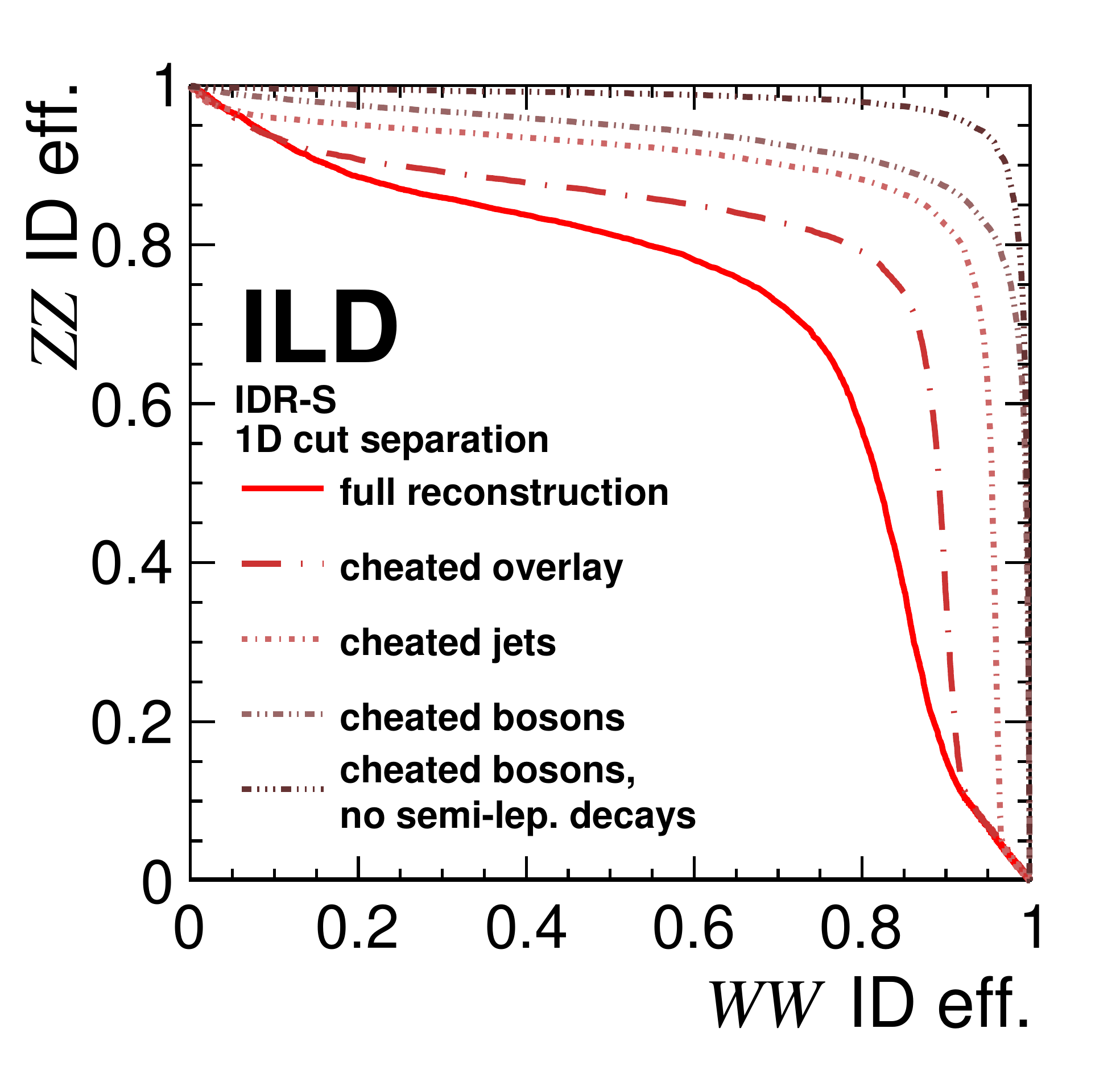}
      \caption{}
      \label{SUBFIG:IDRS_sep_curves}
    \end{subfigure}
    
    \begin{subfigure}[t]{0.5\textwidth}
      \centering
      \includegraphics[width=\textwidth]{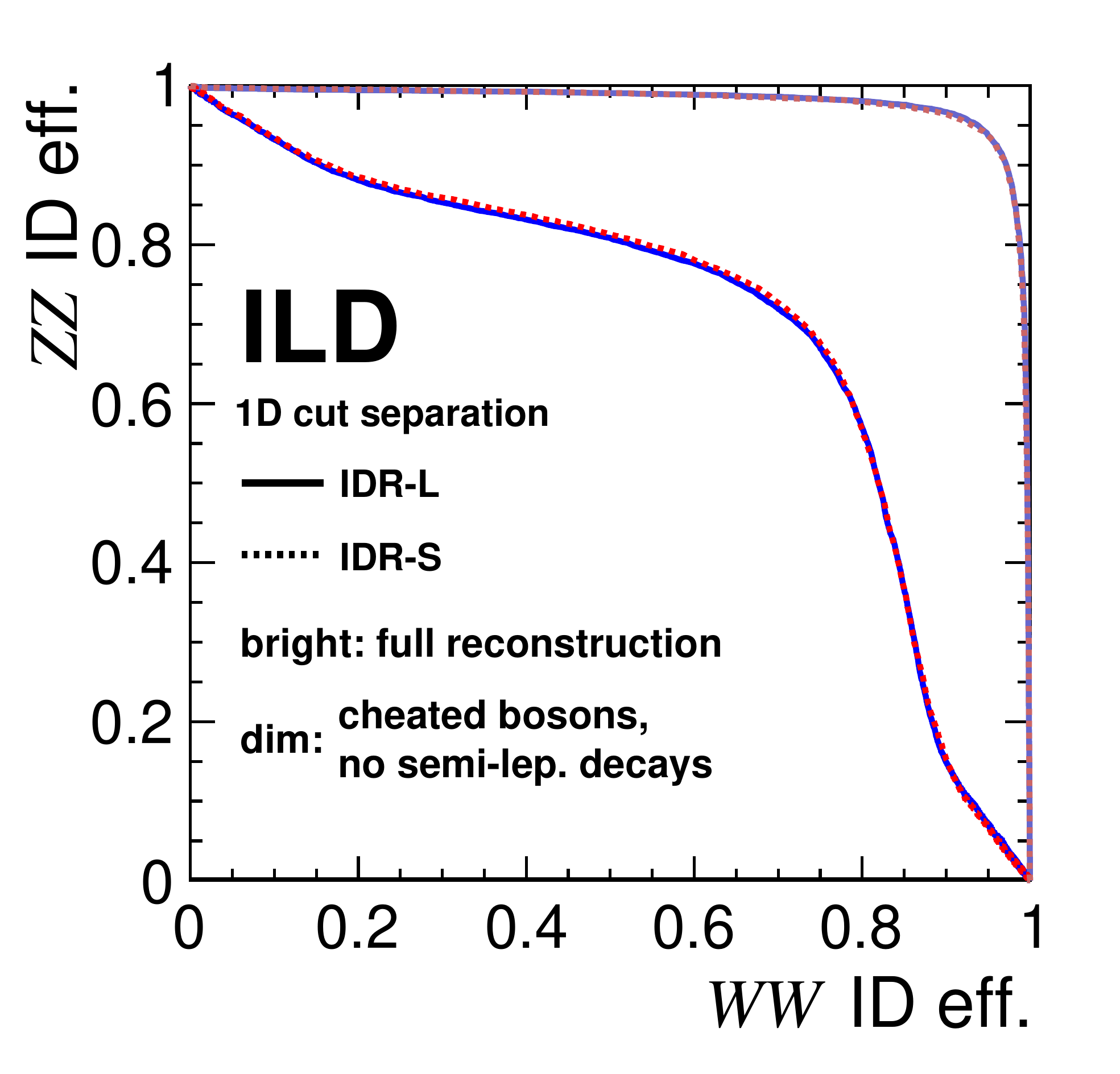}
      \caption{}
      \label{SUBFIG:IDR_sep_curves_ls}
    \end{subfigure}
    \caption{
      Receiver operating characteristic (ROC) curve for a simple $WW/ZZ$ classification cut in the mass average distribution of the two reconstructed vector boson candidates.
      In \subfigref{SUBFIG:IDRL_sep_curves} and \subfigref{SUBFIG:IDRS_sep_curves} for different levels of idealized reconstruction for the large and small ILD models, respectively.
      \subfigref{SUBFIG:IDR_sep_curves_ls} comparing the large and small ILD models at the levels of full high level reconstruction and of fully idealized high level reconstruction.
    }
    \label{FIG:IDRSepCurves}
  \end{figure}
  
  The resulting separation efficiencies for the levels of idealized reconstruction described largely reflect the findings above (tab.~\ref{TAB:WWZZSeparationEfficiencies}).
  Notable is the difference between the full $m_{VV}$ range dataset and the one restricted to $m_{VV}>500\,$GeV.
  The latter shows that on a level of pure detector effects the high-$m_{VV}$ events have a slightly decreased resolution.
  Due to the changed dominant high-level reconstruction effects however, they end up with a slightly better overall resolution after the full reconstruction.
  
  \begin{table}
    \centering
    \caption{
      Efficiencies of a simple $WW/ZZ$ classification cut in the mass average distribution of the two reconstructed vector boson candidates in the large and small ILD models at different levels of idealized high level reconstruction.
      Numbers for the dataset that restricts $m_{VV}>500\,$GeV on generator level are shown as well.
      The cut is choosen to give the same efficiencies for $WW$ and $ZZ$ identification so that only one number is displayed.
    } 
    \label{TAB:WWZZSeparationEfficiencies}
    \begin{tabular}{|l|l|l|l|l|} \hline
      Level & \multicolumn{4}{c|}{$\epsilon_{WW/ZZ} [\%]$} \\ \cline{2-5}
       & \multicolumn{2}{c|}{full $m_{VV}$ range} & \multicolumn{2}{c|}{$m_{VV}>500\,$GeV} \\ \cline{2-5}
       & IDR-L & IDR-S & IDR-L & IDR-S \\ \hline \hline
      Full reconstruction  & 71.1 & 71.5 & 73.0 & 72.9 \\ \hline
      Cheated overlay & 79.6 & 79.4 & 84.6 & 84.0 \\ \hline
      Cheated jets & 86.3 & 85.9 & 86.2 & 85.6 \\ \hline
      Cheated bosons & 88.4 & 88.1 & 86.6 & 86.1 \\ \hline
      No semi-leptonic events & 94.4 & 94.3 & 92.6 & 92.5 \\ \hline
    \end{tabular}
  \end{table}

\section{Correcting semi-leptonic hadron decays\normalsize\protect\footnote{The source code used for the analysis in this section can be found in~\cite{GitHubNuCorrectionRepo}.}}
\begin{figure} \centering
  \includegraphics[scale=1.0]{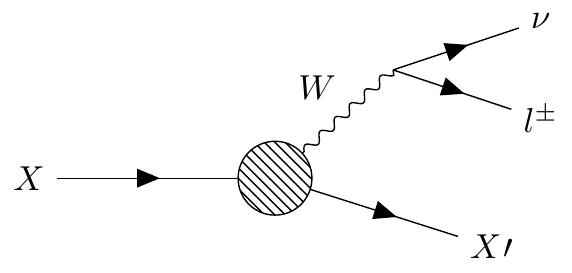}
  \caption{Schematic of a semi-leptonic decay of a hadron.}
  \label{FEY:SemileptonicDecay}
\end{figure}

As described in the previous section, the reconstruction of signal events is in part influenced by semi-leptonic decays in jets.
These decays occur when a heavy hadron of the jet either decays to or radiates a $W$ boson which decays to charged lepton - neutrino pair (fig.~\ref{FEY:SemileptonicDecay}).
Due to the neutrinos in such decays jets can contain some missing energy.
This effect is expected to be stronger in jets originating from heavier quarks (e.g. $b$ and $c$) since such jets contain a higher number of heavy hadrons.

\begin{figure}
  \centering
  \includegraphics[width=0.5\textwidth]{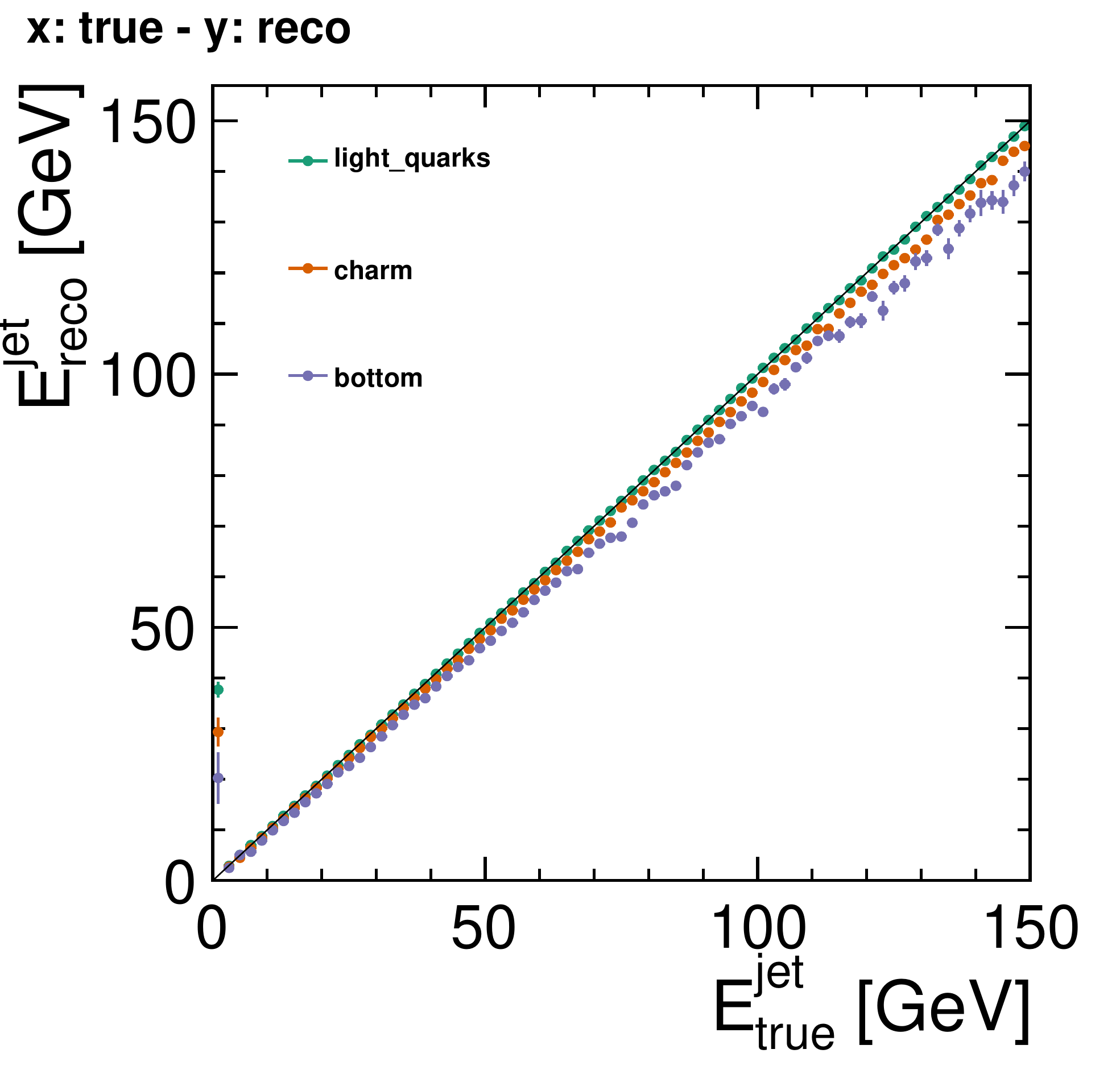}
  \caption[
  Jet energy scale for individual jets of different flavour origins found by \texttt{TrueJet}.
  ]{%
  Jet energy scale for individual jets in signal events. 
  Individual jets are found using \texttt{TrueJet}. 
  Colours represent different initial quark origins of the jets.
  }
  \label{FIG:JES_flavour_dependent}
\end{figure}

Such a dependency on the jet flavour can be seen directly in the jet energy scale (JES) of the signal event jets separated by their generator level quark flavour (fig.~\ref{FIG:JES_flavour_dependent}).
Light quark jets show a diagonal JES\footnote{Here and in the following the $0\,$GeV bin shows anomalous behavior likely to be caused by faulty behavior in \texttt{TrueJet}.} as expected because they contain only a small of heavy hadrons produced in gluon radiation.
Jets originating from heavy quarks show a visible tilt of the JES.
The effect is stronger for $b$ jets since they contain both $B$ and $C$ hadrons while $c$ jets contain only $C$ mesons.

Because roughly $30\%$ of $W$ and $Z$ decays contain $c$ or $b$ quarks these semi-leptonic decays will influence many electro-weak analyses~\cite{PhysRevD.98.030001}.
Here, two approaches to correct for such decays are studied on a proof-of-principle level for signal events in the IDR-L model.

\subsection{Spectrum-based correction}

\begin{figure}
    \centering
    \includegraphics[width=0.5\textwidth, trim={0 0 5cm 0},clip]{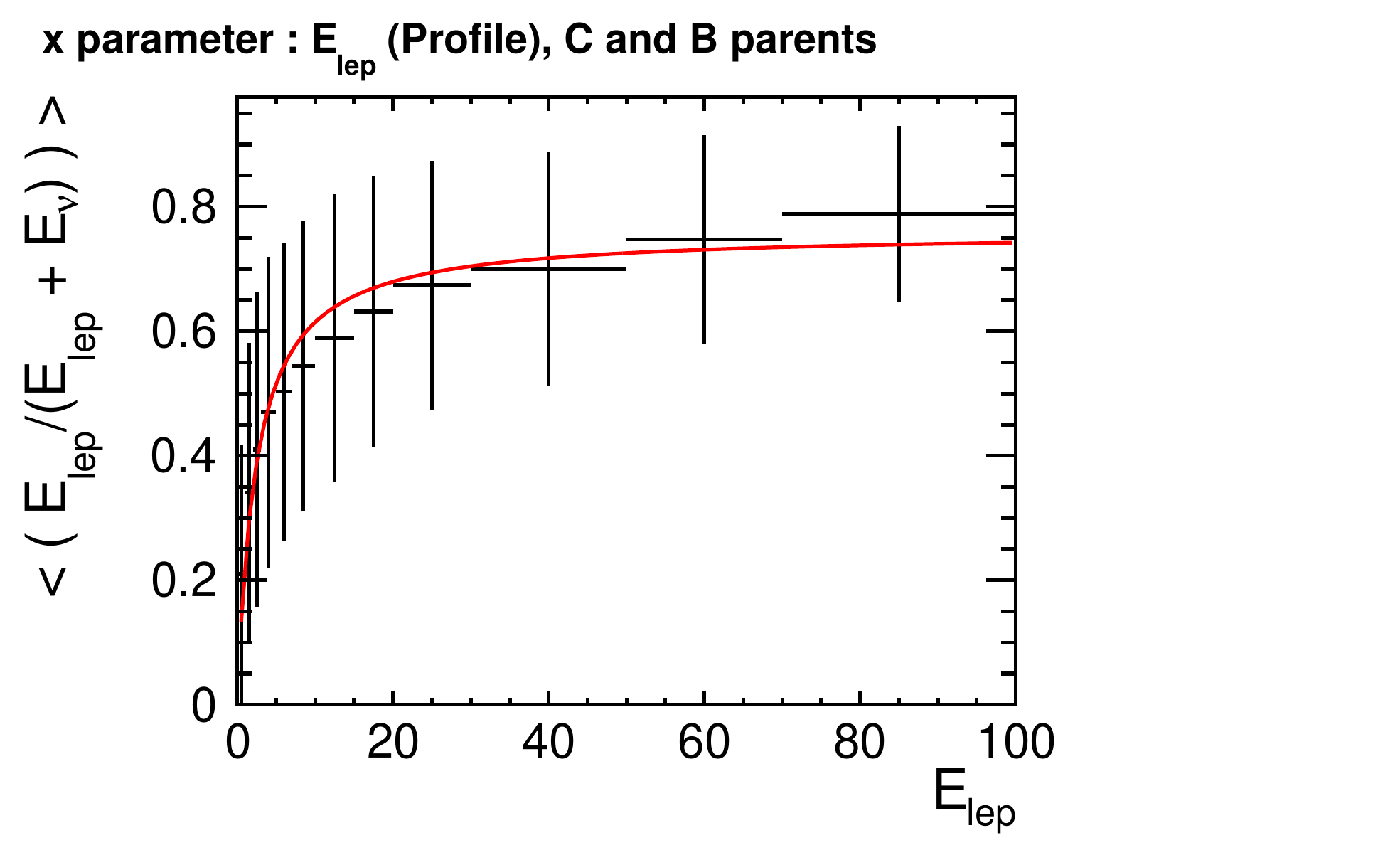}
    \caption{
      Average charged lepton energy fraction in the charged lepton - neutrino pair of semi-leptonic $B$ and $C$ hadron decays in signal events. 
      The red curve shows the fit described in eq.~\ref{EQ:RoughNuCorr_x_fit}.
    }
    \label{FIG:RoughNuCorr_x_fit}
  \end{figure}

  This first approach is based on the correlation of the neutrino energy $E_{\nu}$ and the charged lepton energy $E_{l^{\pm}}$ in semi-leptonic decays.
  For this the fraction of the charged lepton energy $x$ is introduced 
  \begin{equation}
    x = \frac{E_{l^{\pm}}}{E_{l^{\pm}} + E_{\nu}}
  \end{equation}
  and a fit is performed to gain $<x>$ as a function only of the charged lepton energy
  \begin{equation}\label{EQ:RoughNuCorr_x_fit}
     <x> \left( E_{l^{\pm}} \right) = a \cdot \frac{E_{l^{\pm}}}{E_{l^{\pm}} + b} \, .
  \end{equation}
  The fit is performed on the generator level energies of all semi-leptonic $C$ and $B$ decays of the signal events and returns fit parameters of $a = 0.76$ and $b = 2.4$ (fig.\ref{FIG:RoughNuCorr_x_fit}). 
  With this fit a correction of the jet energy solely based on the measured charged lepton is possible
  \begin{equation} \label{EQ:RoughNuCorr_x_corr}
    E_{jet}^{corr} = E_{jet}^{meas} + \left( \frac{1}{<x>} - 1 \right) E_{l^{\pm}} \, .
  \end{equation}
  
\begin{figure}
    \centering
    \begin{subfigure}[t]{0.5\textwidth}
      \centering
      \includegraphics[width=\textwidth]{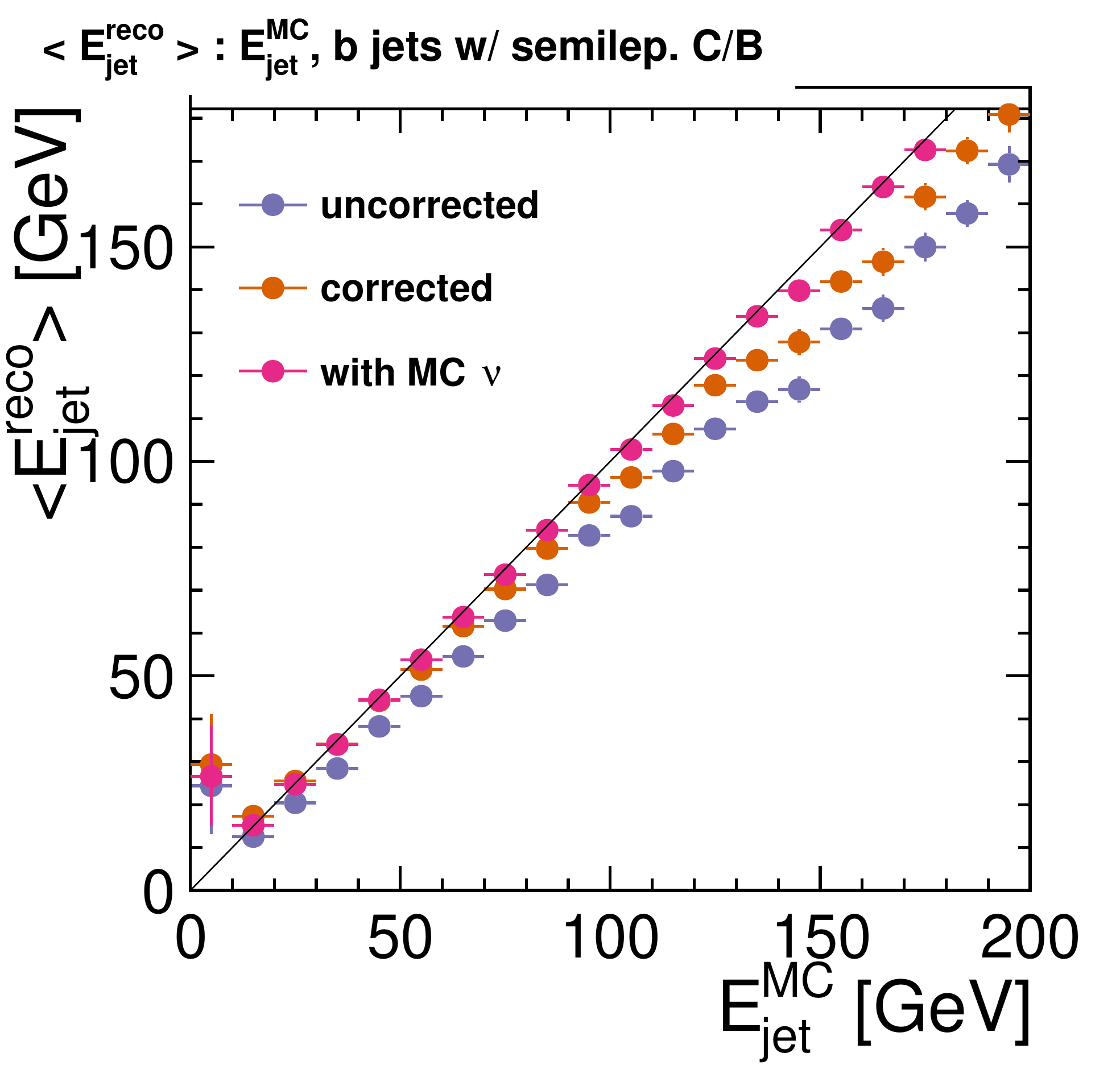}
      \caption{}
      \label{SUBFIG:RoughNuCorr_onbTJ_onlysemilep}
    \end{subfigure}%
    \begin{subfigure}[t]{0.5\textwidth}
      \centering
      \includegraphics[width=\textwidth]{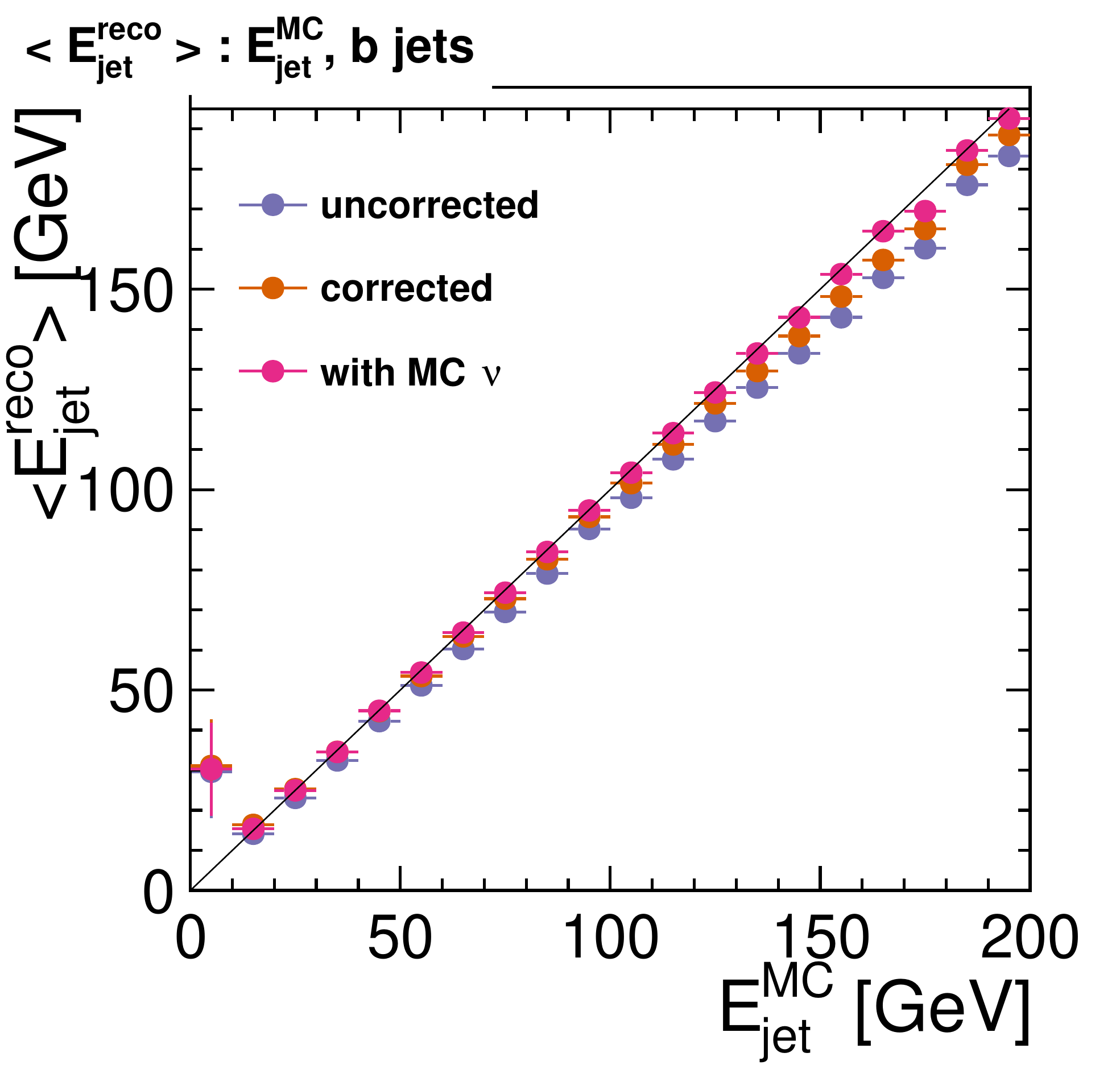}
      \caption{}
      \label{SUBFIG:RoughNuCorr_onbTJ_all}
    \end{subfigure}
    
    \caption{%
    Jet energy scale for test of average-based neutrino correction (eq. \ref{EQ:RoughNuCorr_x_corr}) on individual \texttt{TrueJet} jets. 
    The correction is applied to generator level charged lepton-neutrino vertices and uses the generator level charged lepton energy. 
    JES is shown without correction, with correction and with added generator level neutrino energies.\\
    \subfigref{SUBFIG:RoughNuCorr_onbTJ_onlysemilep} Only $b$ jets which contain semi-leptonic decays are used.\\
    \subfigref{SUBFIG:RoughNuCorr_onbTJ_all} All $b$ jets are shown, the correction is only applied when semi-leptonic decays are found on generator level.\\
    }
    \label{FIG:RoughNuCorr_onbTJ}
  \end{figure}

  A test of this correction is performed on the JES of the $b$ jets in the signal events as determined by \texttt{TrueJet} (fig.\ref{FIG:RoughNuCorr_onbTJ}).
  All jet reconstruction aspects except for the correction of semi-leptonic decays are idealized to observe the influence of the correction in isolation.
  The correction uses the generator level charged leptons from $C$ or $B$ hadron decays in the generator level jet.
  In this proof-of-principle check the correction performs a visible tilt of the JES of the targeted jets towards the diagonal.
  
  While this result is promising future studies will need to show how well the correction can perform when only using detector level information, especially for tagging of $b$ and $c$ jets.  
  Furthermore the effect on the jet energy resolution is not studied here and will be the determining factor for the usability of any correction.

\subsection{Exact reconstruction correction}
    
\begin{figure}
    \centering
    \begin{subfigure}[t]{0.5\textwidth}
      \centering
      \includegraphics[width=\textwidth]{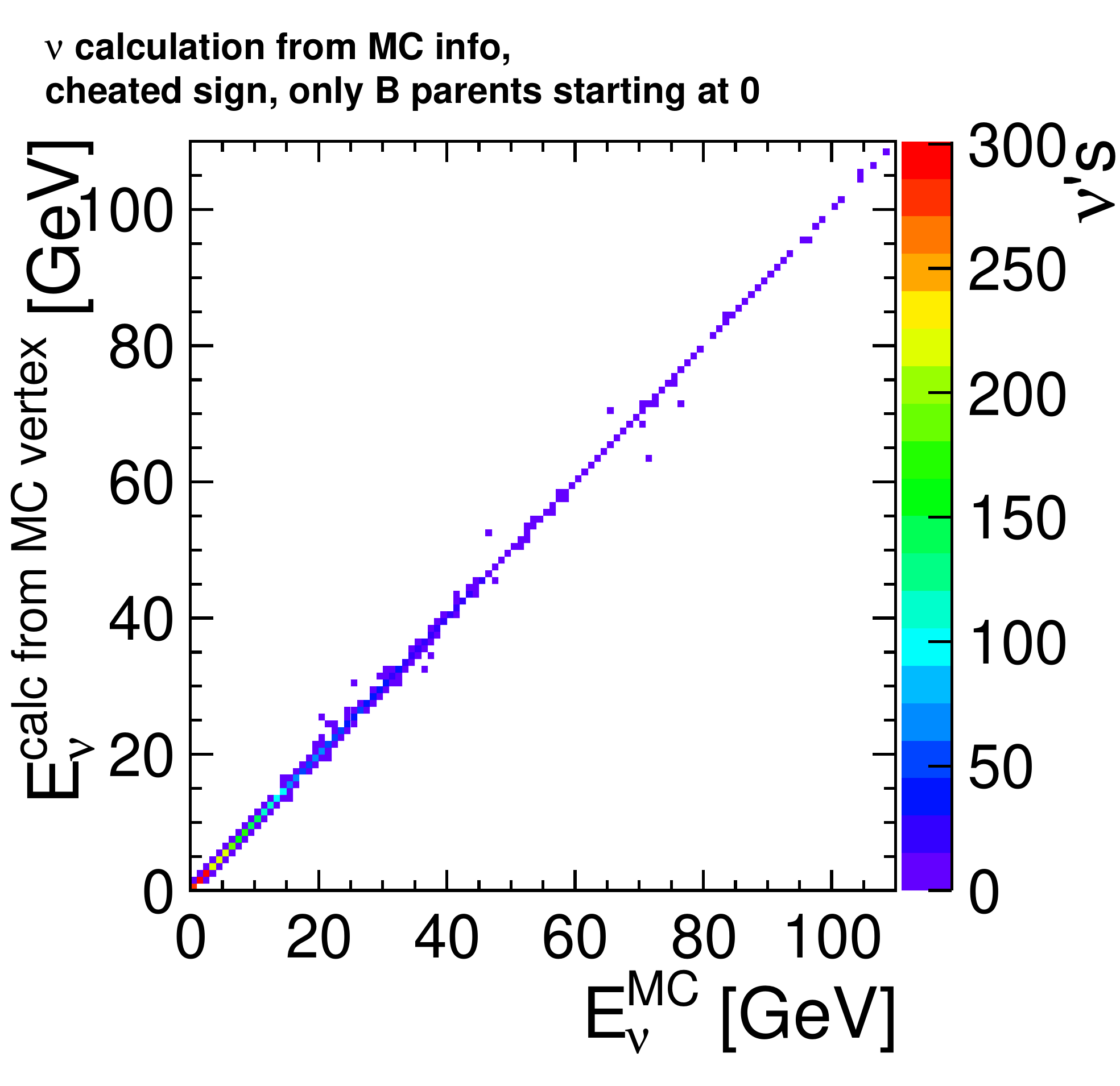}
      \caption{}
      \label{SUBFIG:NuReconstruction_allcheated_toHighE}
      \end{subfigure}%
      \begin{subfigure}[t]{0.5\textwidth}
        \centering
        \includegraphics[width=\textwidth]{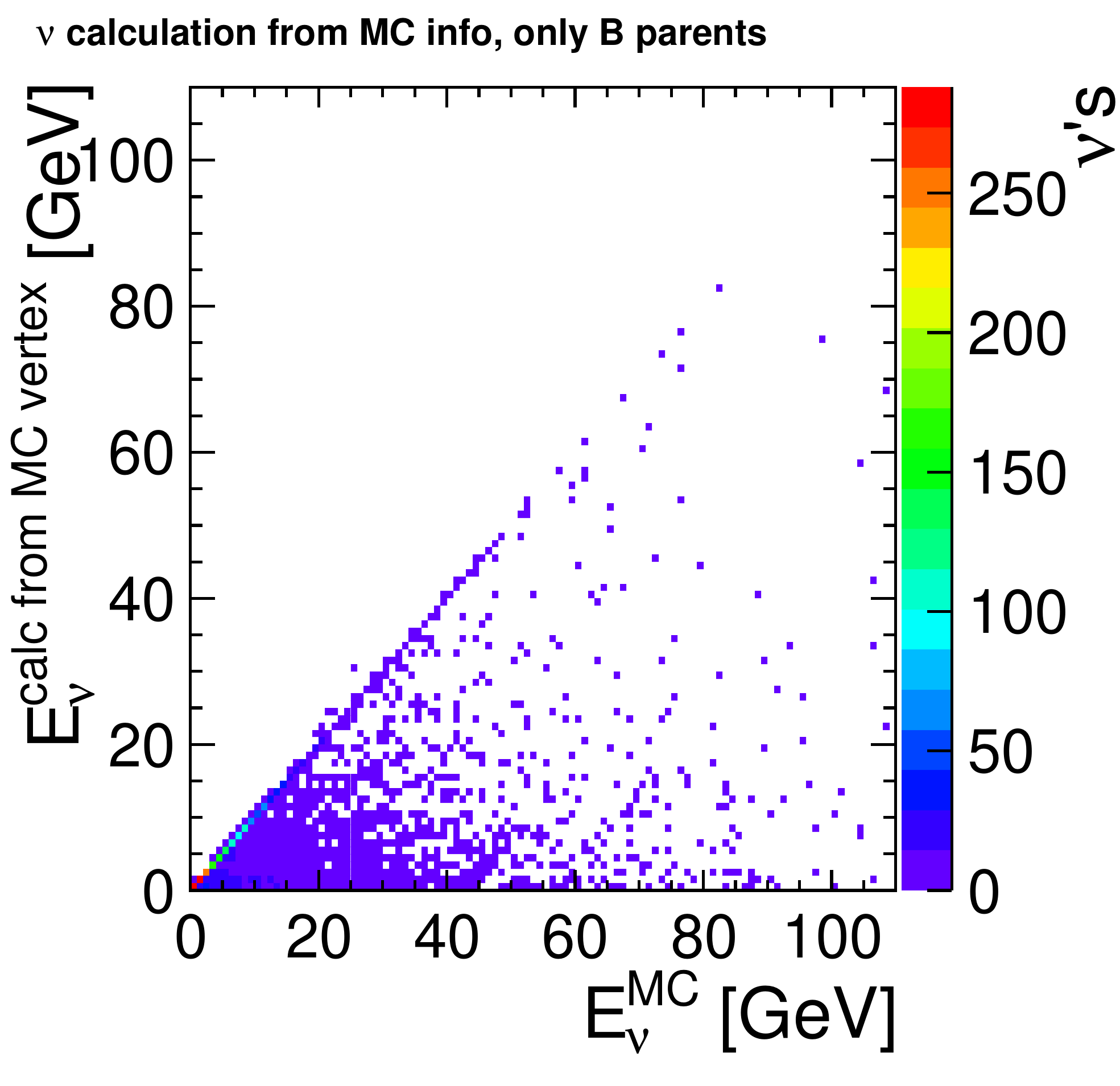}
        \caption{}
        \label{SUBFIG:NuReconstruction_minus_toHighE}
      \end{subfigure}
      
      \begin{subfigure}[t]{0.5\textwidth}
        \centering
        \includegraphics[width=\textwidth]{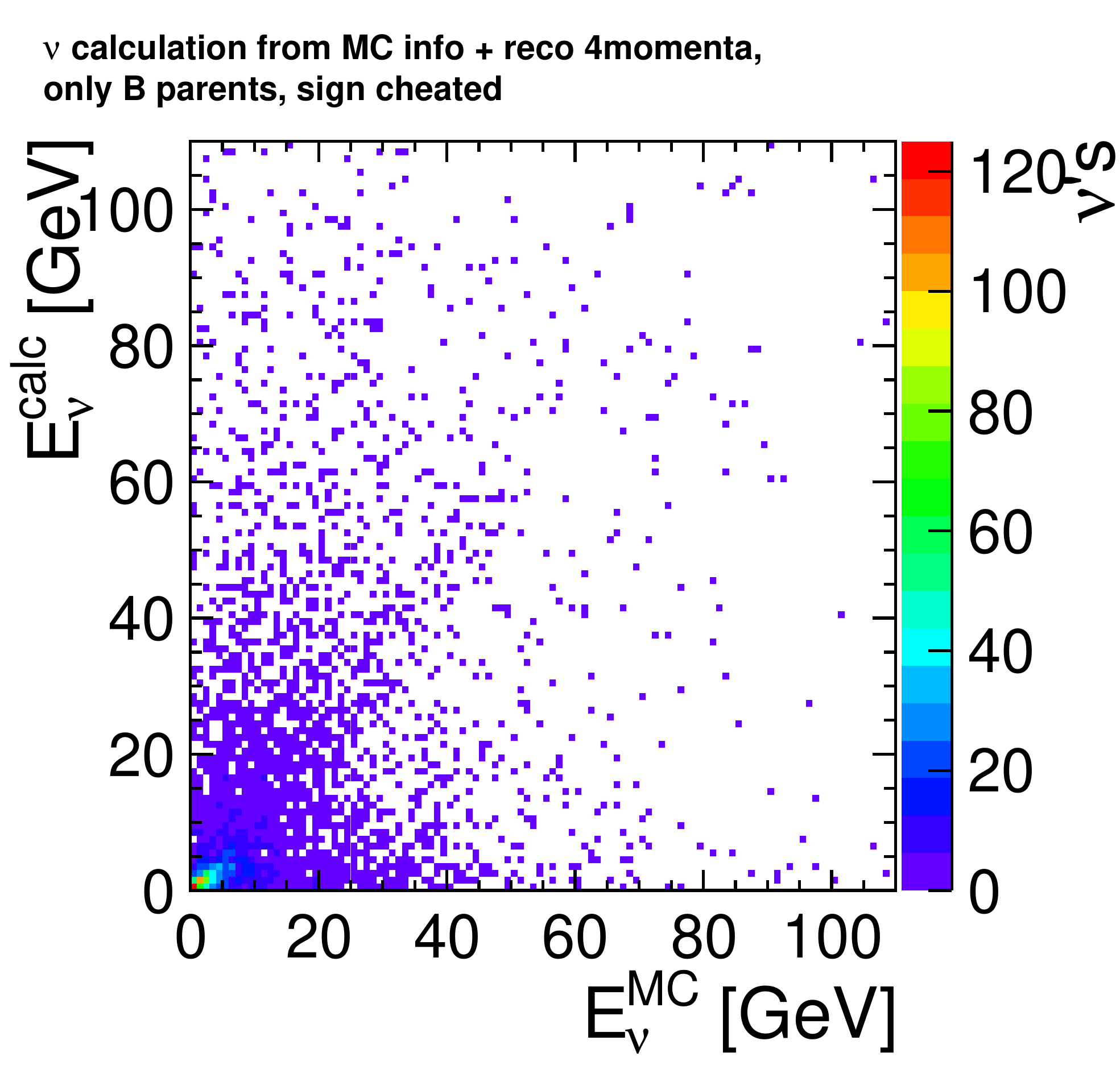}
        \caption{}
        \label{SUBFIG:NuReconstruction_recomom_signcheated_toHighE}
    \end{subfigure}%
    \begin{subfigure}[t]{0.5\textwidth}
      \centering
      \includegraphics[width=\textwidth]{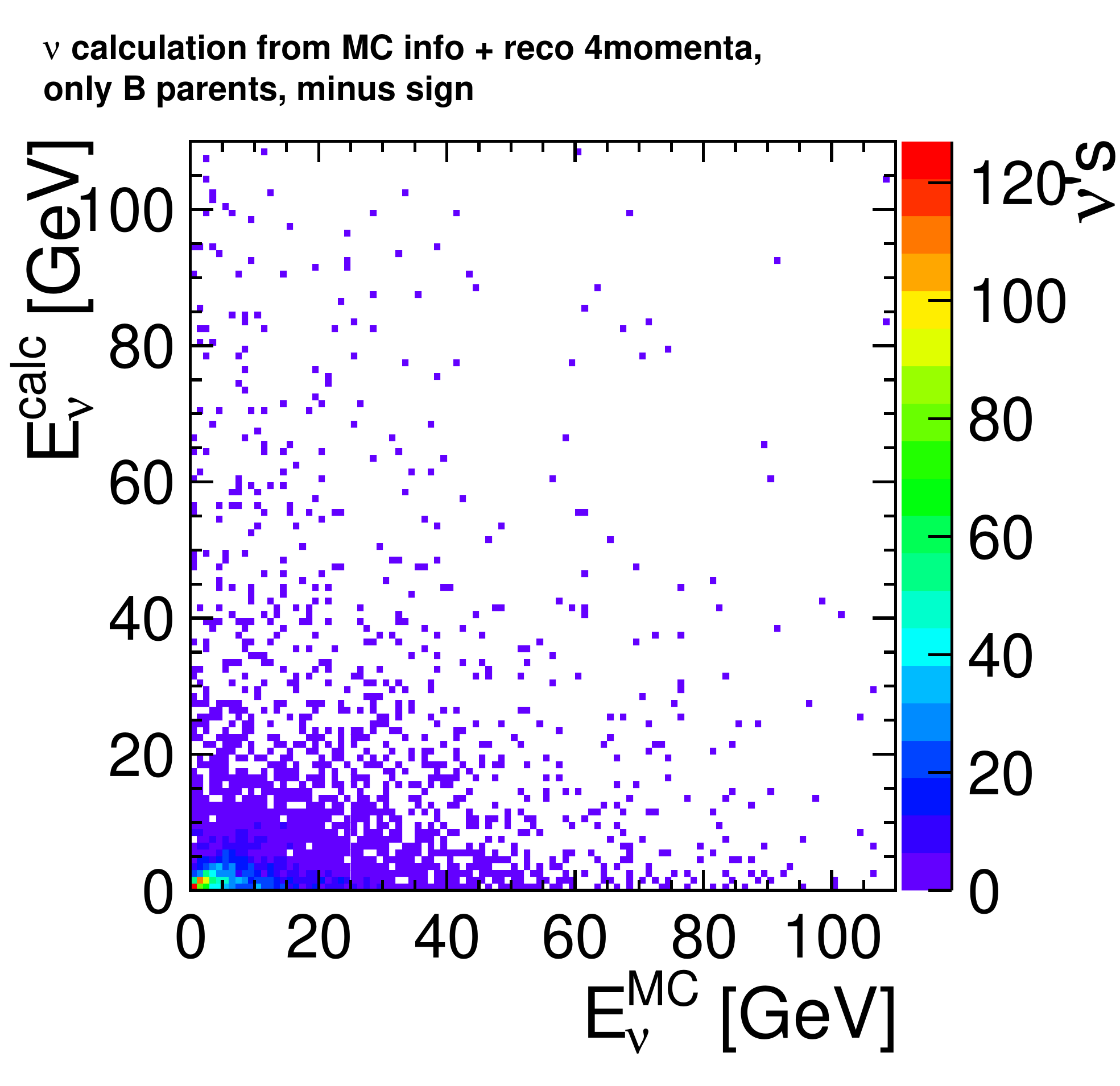}
      \caption{}
      \label{SUBFIG:NuReconstruction_withrecomom_minus_toHighE}
    \end{subfigure}
    \caption[
    Distribution of reconstructed versus true neutrino energy for fest of kinematic neutrino reconstruction.
    ]{%
    Distribution of calculated neutrino energy depending on the true neutrino energy for test of neutrino reconstruction in semi-leptonic decays according to eq. \eqref{EQ:ExactNeutrinoSolution}. 
    In all plots the mass and direction of the initial hadron are taken from generator level information.\\
    \subfigref{SUBFIG:NuReconstruction_allcheated_toHighE} All input variables taken from generator level information of the decay chain and the correct solution is chosen using the true neutrino momentum. \\
    \subfigref{SUBFIG:NuReconstruction_minus_toHighE} Input variables taken from generator level as well but minus-sign solution of eq. \eqref{EQ:ExactNeutrinoSolution} is chosen. \\
    \subfigref{SUBFIG:NuReconstruction_recomom_signcheated_toHighE} Using the solution which gives the correct answer when generator level information is used, but calculating the neutrino momentum using the reconstructed visible 4-momentum. \\
    \subfigref{SUBFIG:NuReconstruction_withrecomom_minus_toHighE} Reconstructed 4-momenta of the visible particles are used and the minus sign of eq. \eqref{EQ:ExactNeutrinoSolution} is chosen.
    }
    \label{FIG:NuReconstruction_toHighE}
  \end{figure}

  An exact correction of the neutrino four-momentum may be possible from four-momentum conservation if the decay vertex and the decay product are measured accurately, and if an assumption can be made about the parents mass.
  The neutrino momentum perpendicular to the direction of the decaying hadron $p_{\nu,\perp}$ can be easily inferred from the visible decay products if the hadrons flight direction is known
  \begin{equation}
    p_{\nu,\perp} = - p_{\text{vis},\perp} \,.
  \end{equation}
  However, the formula for the neutrino momentum parallel to the parent direction has two solutions \footnote{For the full derivation and an explanation of the assumptions see \cite{Beyer:416692}, Appendix C.}:
  \begin{equation} \label{EQ:ExactNeutrinoSolution}
    p_{\nu,\parallel} = \frac{1}{2 \cdot D} \cdot \left(-A \, \pm\, \sqrt{A^2 - BD} \right) \, 
  \end{equation}
  where
   \begin{align}
    A & =p_{\text{vis},\parallel} \cdot ( 2 p_{\text{vis},\perp}^2 + m_{\text{vis}}^2 - m_{X}^2) \\
    B & =4 p_{\text{vis},\perp}^2 \cdot E_{\text{vis}}^2 - ( 2 p_{\text{vis},\perp}^2 + m_{\text{vis}}^2 - m_{X}^2 )^2 \\
    D & =E_{\text{vis}}^2- p_{\text{vis},\parallel}^2
  \end{align}
  with $vis$ marking kinematics of the visible decay products and $X$ marking those of the decaying hadron.
  The correct solution is not known from pure kinematics and needs to be derived using further information.\\
  Further, due to numerical uncertainties the square-root in eq.\ \ref{EQ:ExactNeutrinoSolution} may become imaginary.
  Here, such cases are treated by setting the square-root to zero.
  In a proof-of-principle test of this new approach which is based solely on generator level information this leads to minor off-diagonal effects (see fig.\ \ref{SUBFIG:NuReconstruction_allcheated_toHighE}).
  When choosing only the negative sign solution of eq.\ \ref{EQ:ExactNeutrinoSolution} to test the influence of the unknown sign a large number of reconstructed neutrino momenta remains diagonal (see fig.\ \ref{SUBFIG:NuReconstruction_minus_toHighE}). 
  However, using reconstructed four-momenta in the calculation strongly smears the calculated neutrino four-momenta (see fig.\ \ref{SUBFIG:NuReconstruction_recomom_signcheated_toHighE}).
  Whether this limits the applicability of this approach remains to be tested.
  
  $ $ 
      
  These tests largely rely on generator level information.
  Both take the identification of semi-leptonic decay vertices from generator level.
  The exact approach additionally uses knowledge about the parent particle mass and direction.
  
  How this information can instead be extracted on the reconstructed level remains to be tested.
  Additionally, the influence on the analysis is not investigated here.

\section{Conclusion}
In this note an analysis of the Vector Boson Scattering final state $\eP\eM\rightarrow \nu\nubar q\qbar q\qbar$ has been performed in full ILD simulation for a $1\,$TeV ILC.
The influence of individual reconstruction effects is investigated by cheating each effect using generator level information and using the separation of hadronic $WW$ and $ZZ$ decays as benchmark.
Standard reconstruction techniques based on previous studies are used and have not been further optimized~\cite{Chierici:2001ar}.

In this setup the separation of hadronic $WW$ and $ZZ$ decays is found to be limited by the removal of beam backgrounds, the jet clustering, and semi-leptonic decays within jets. 
Which effect is dominating is dependent on the invariant mass of the di-boson system.
Future studies may improve upon these obstacles using tools like advanced jet clustering algorithms~\cite{Boronat:2014hva}, kinematic fitting~\cite{Beckmann:2010ib}, dedicated beam background removal and corrections for semi-leptonic decays. 

Corrections of semi-leptonic hadron decays have been tested on a proof-of-principal level.
An exact reconstruction of such a decay shows strong sensitivity to the decay product four momentum.
Using a correction which is based on the average kinematic spectrum of these decays may reduce this sensitivity while still improving the event reconstruction.
Studies towards systematically applying such corrections on detector level are currently ongoing.

A full aQGC sensitivity analysis has not been performed here.
However, studies have been performed on generator level while assuming some general detector effects~\cite{Fleper:2016frz}.
These assumptions seem achievable with ILD and similar limits can be expected.

\section{Acknowledgments}
We would like to thank the LCC generator working group and the ILD software working group for providing the simulation and reconstruction tools and producing the Monte Carlo samples used in this study.
This work has benefited from computing services provided by the ILC Virtual Organization, supported by the national resource providers of the EGI Federation and the Open Science GRID.

\clearpage

\section{References}
\customprintbibliography

\clearpage
\appendix
\setappendixnumbering

\section{Appendix}
\subsection{High-$\boldsymbol{m_{VV}}$ dataset distributions}

\begin{figure}
  \centering
  \begin{subfigure}[t]{0.5\textwidth}
    \centering
    \includegraphics[width=\textwidth]{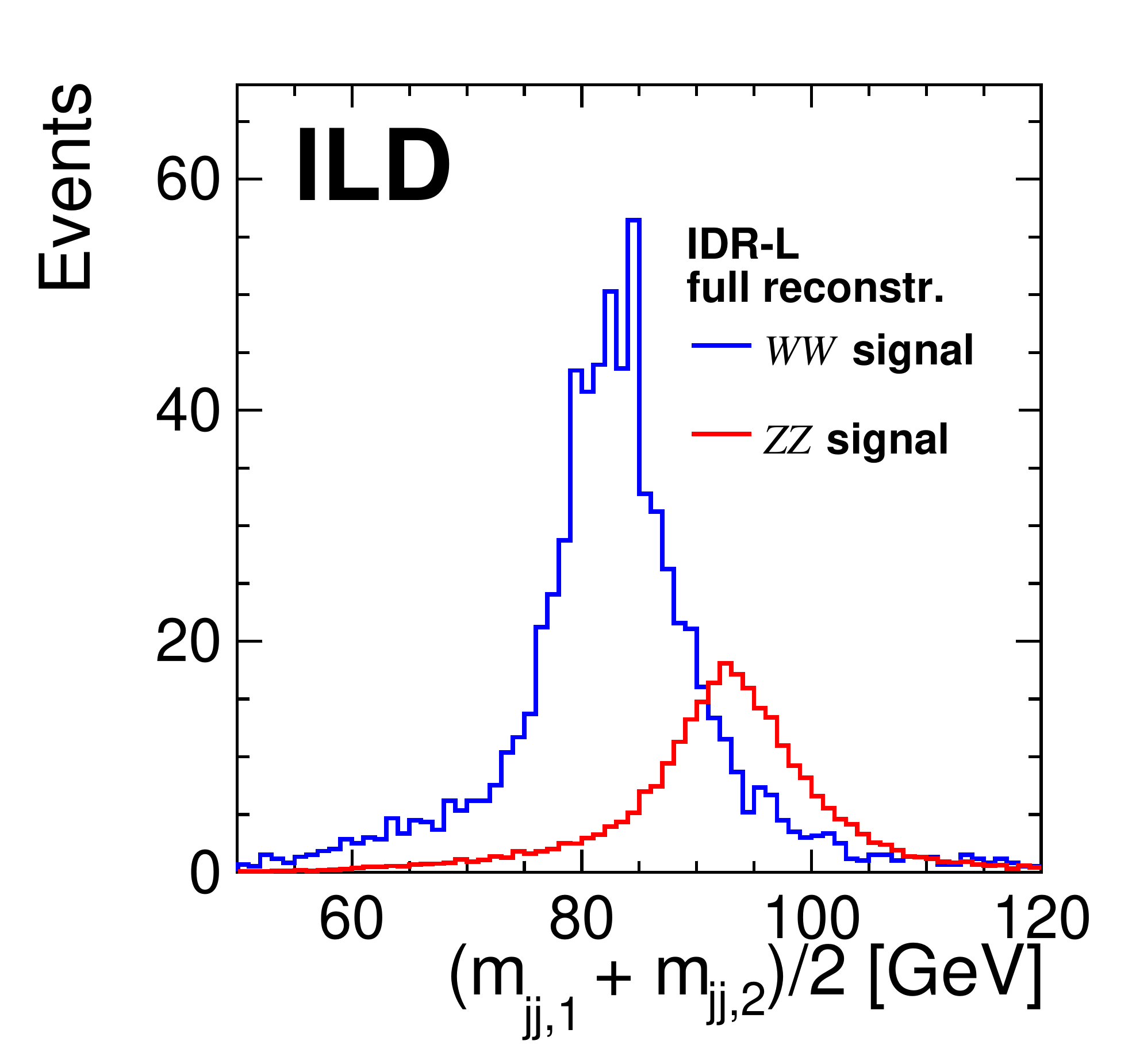}
    \caption{}
    \label{SUBFIG:IDRL_m_highQ2}
  \end{subfigure}%
  \begin{subfigure}[t]{0.5\textwidth}
    \centering
    \includegraphics[width=\textwidth]{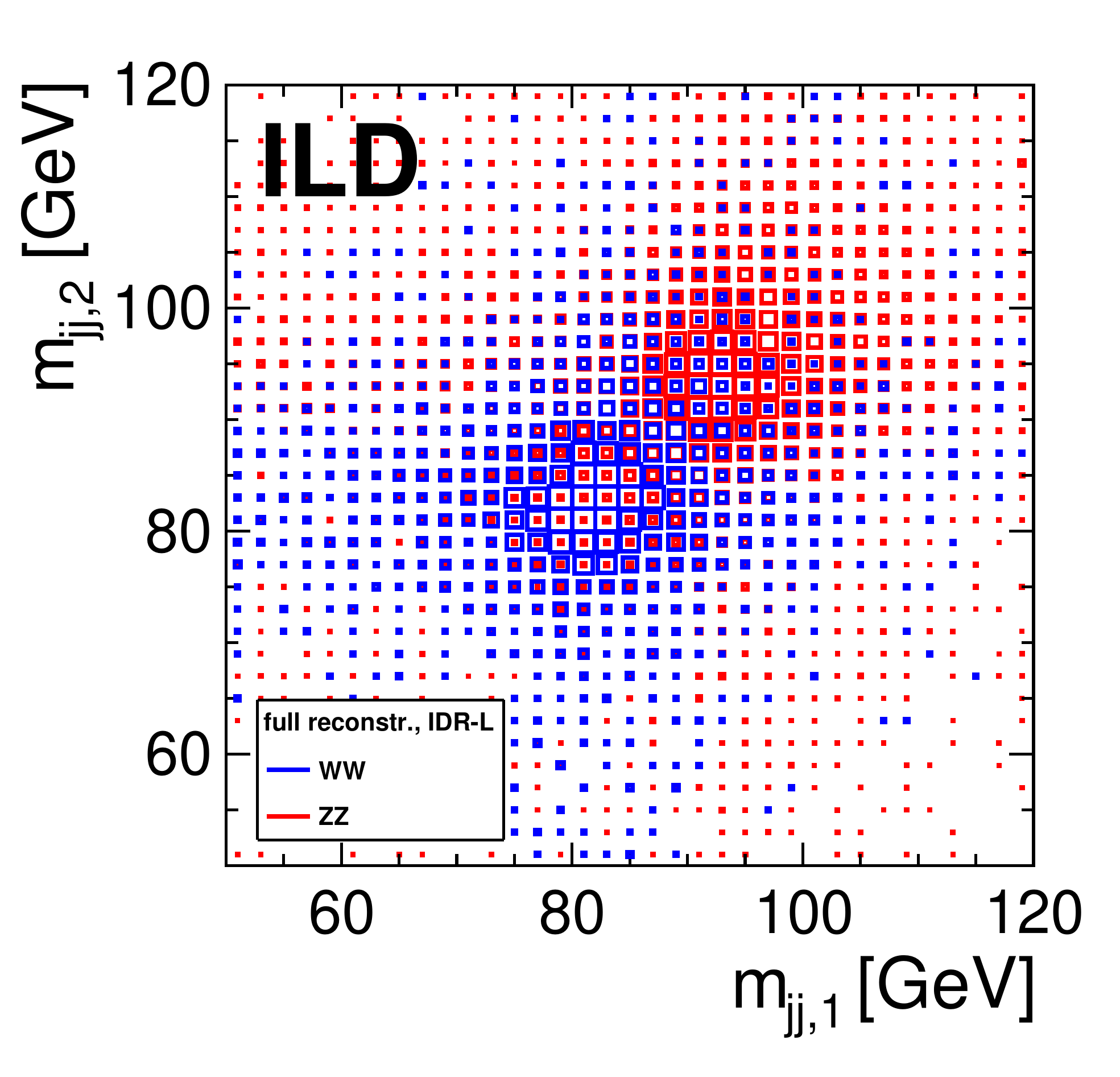}
    \caption{}
    \label{SUBFIG:IDRL_m_m_highQ2}
  \end{subfigure}
  
  \begin{subfigure}[t]{0.5\textwidth}
    \centering
    \includegraphics[width=\textwidth]{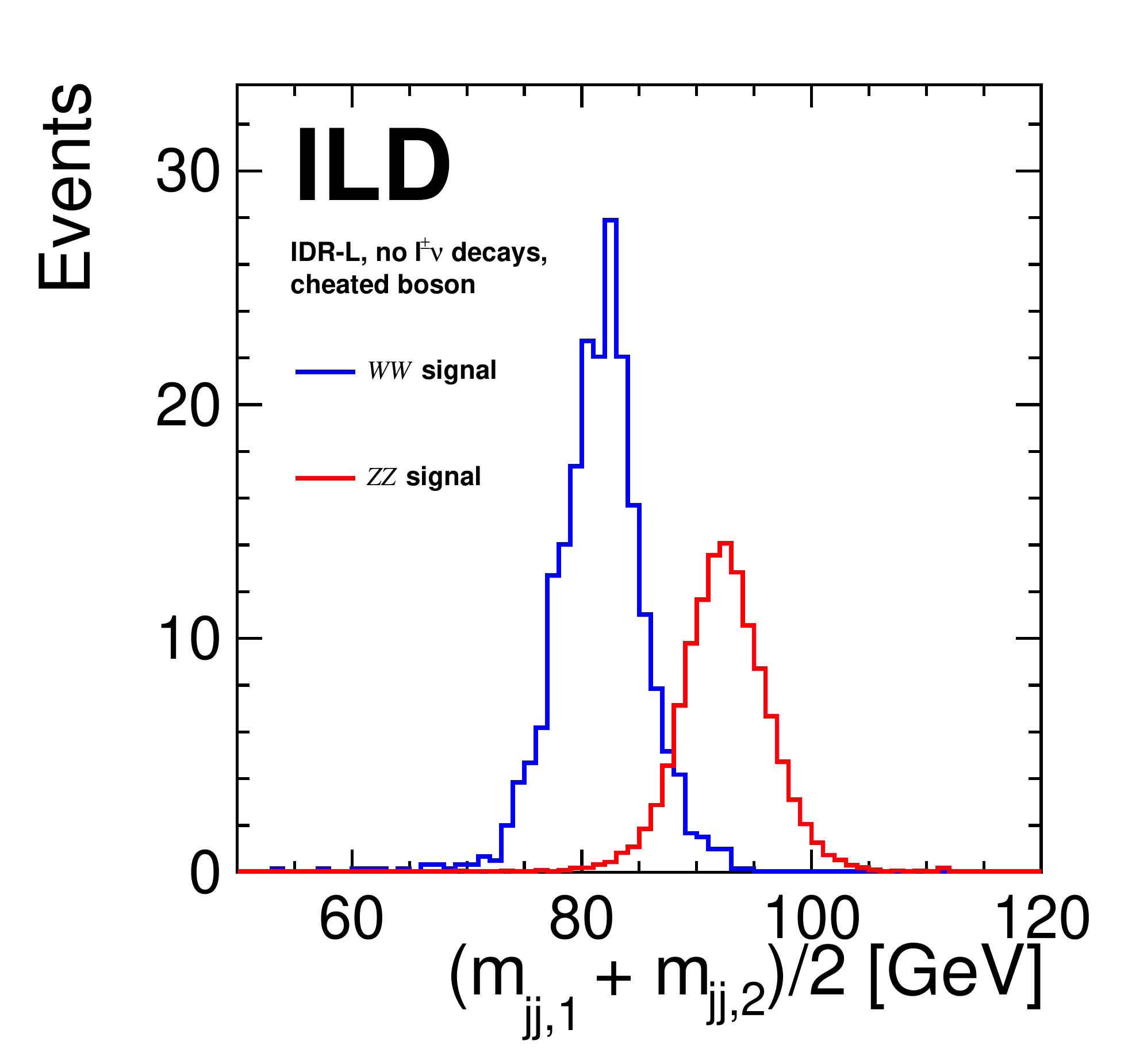}
    \caption{}
    \label{SUBFIG:IDRL_m_icn_noSLD_highQ2}
    \end{subfigure}%
    \begin{subfigure}[t]{0.5\textwidth}
      \centering
      \includegraphics[width=\textwidth]{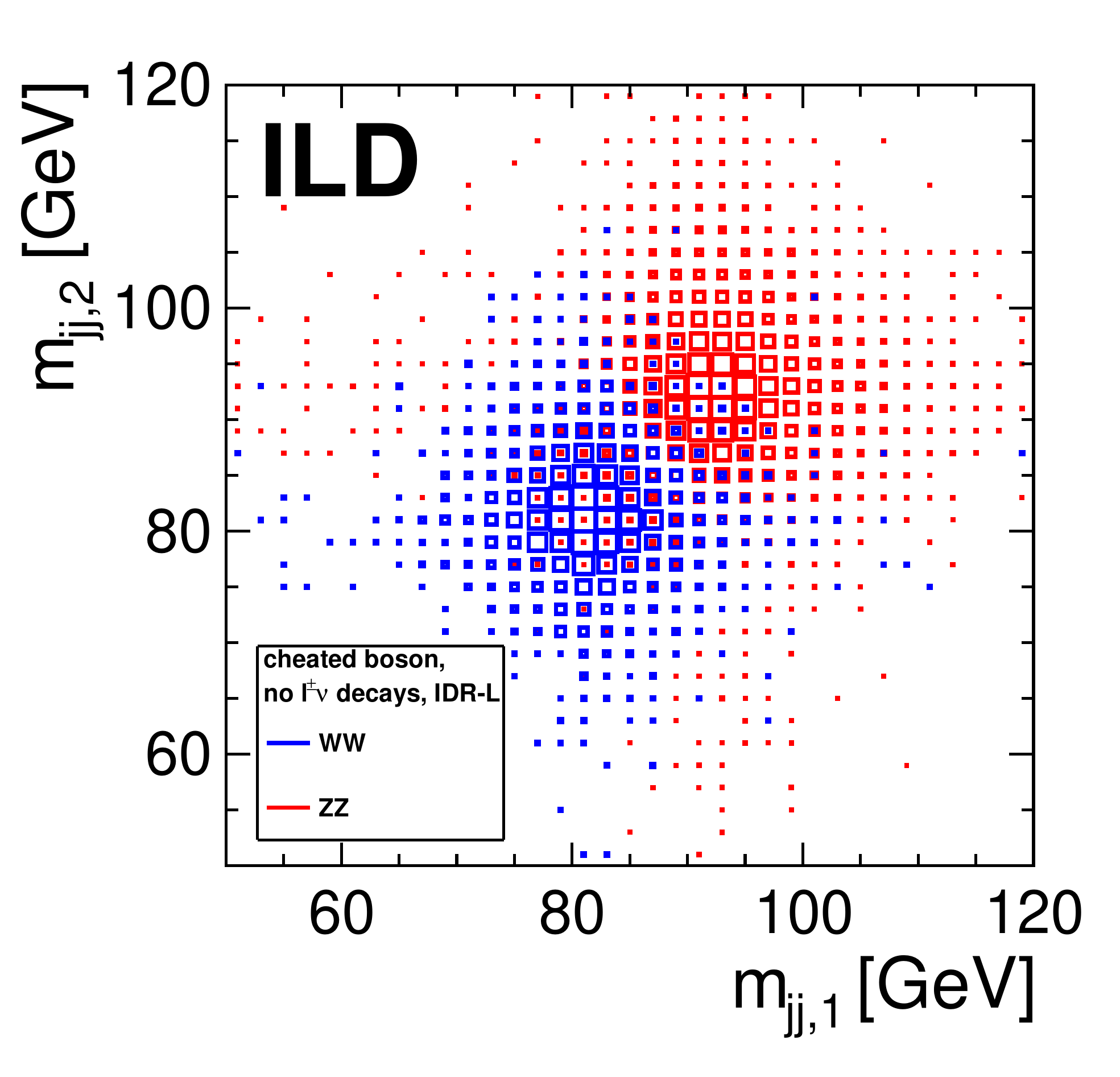}
      \caption{}
      \label{SUBFIG:IDRL_m_m_icn_noSLD_highQ2}
    \end{subfigure}%
  \caption{
    Reconstructed vector boson candidate mass spectra (\protect\subref{SUBFIG:IDRL_m_highQ2},\protect\subref{SUBFIG:IDRL_m_m_highQ2}) from full simulation and (\protect\subref{SUBFIG:IDRL_m_icn_noSLD_highQ2},\protect\subref{SUBFIG:IDRL_m_m_icn_noSLD_highQ2}) with idealized high level reconstruction and only for events without semi-leptonic decays in jets.
    The dataset with a $m_{VV}>500\,$GeV restriction on generator level is used with the large ILD model.
    (\protect\subref{SUBFIG:IDRL_m_highQ2},\protect\subref{SUBFIG:IDRL_m_icn_noSLD_highQ2}) Average of both masses, weighted to nominal luminosity of $1\,$ab$^{-1}$ of $1\,$TeV ILC running. 
    (\protect\subref{SUBFIG:IDRL_m_m_highQ2},\protect\subref{SUBFIG:IDRL_m_m_highQ2}) Mass spectrum for both vector boson candidates, with $WW$ and $ZZ$ spectra individually normalized.
  }
  \label{FIG:MassPlotsIDRL_highQ2}
\end{figure}

\begin{figure}
  \centering
  \begin{subfigure}[t]{0.5\textwidth}
    \centering
    \includegraphics[width=\textwidth]{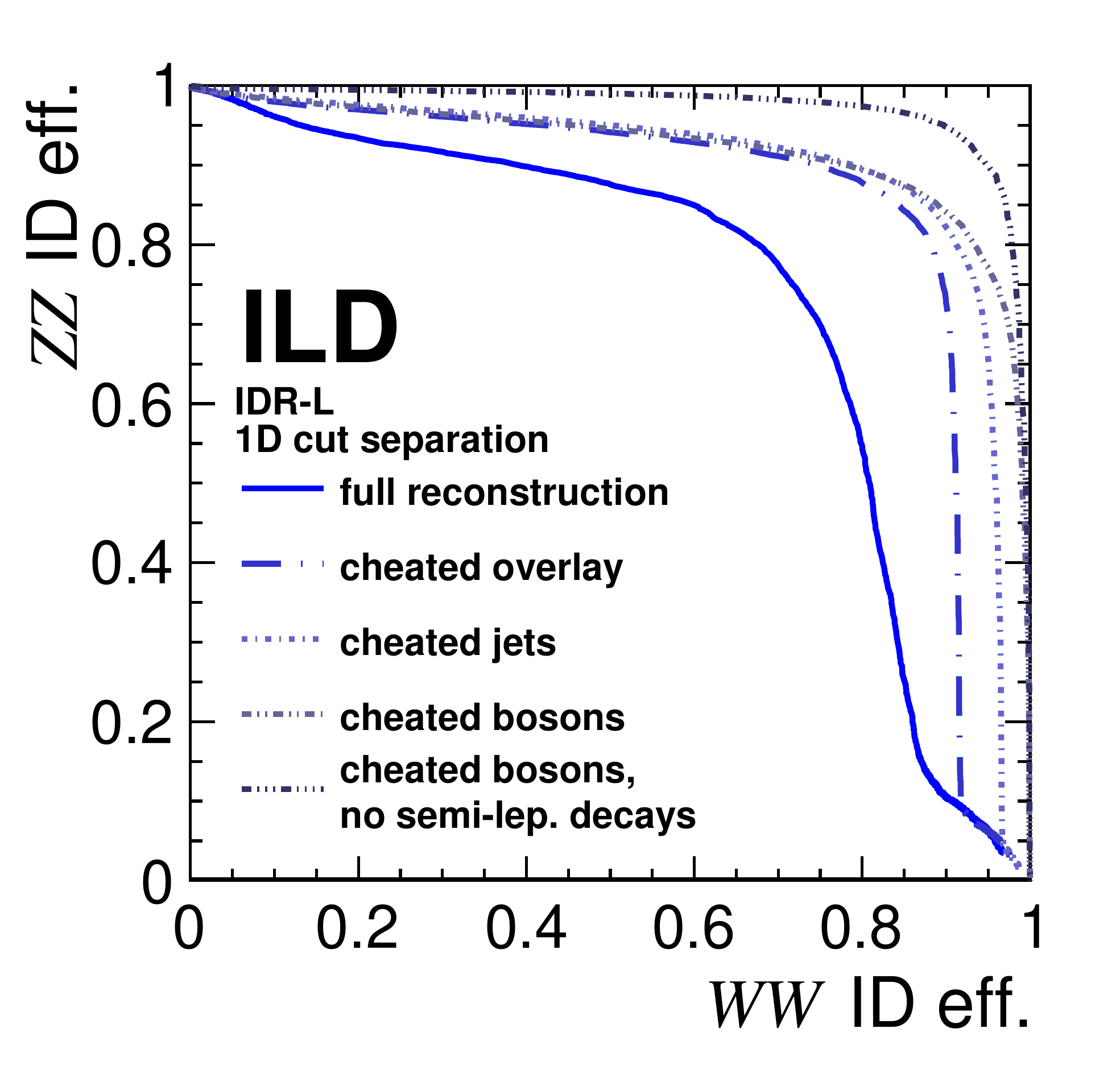}
    \caption{}
    \label{SUBFIG:IDRL_sep_curves_highQ2}
  \end{subfigure}%
  \begin{subfigure}[t]{0.5\textwidth}
    \centering
    \includegraphics[width=\textwidth]{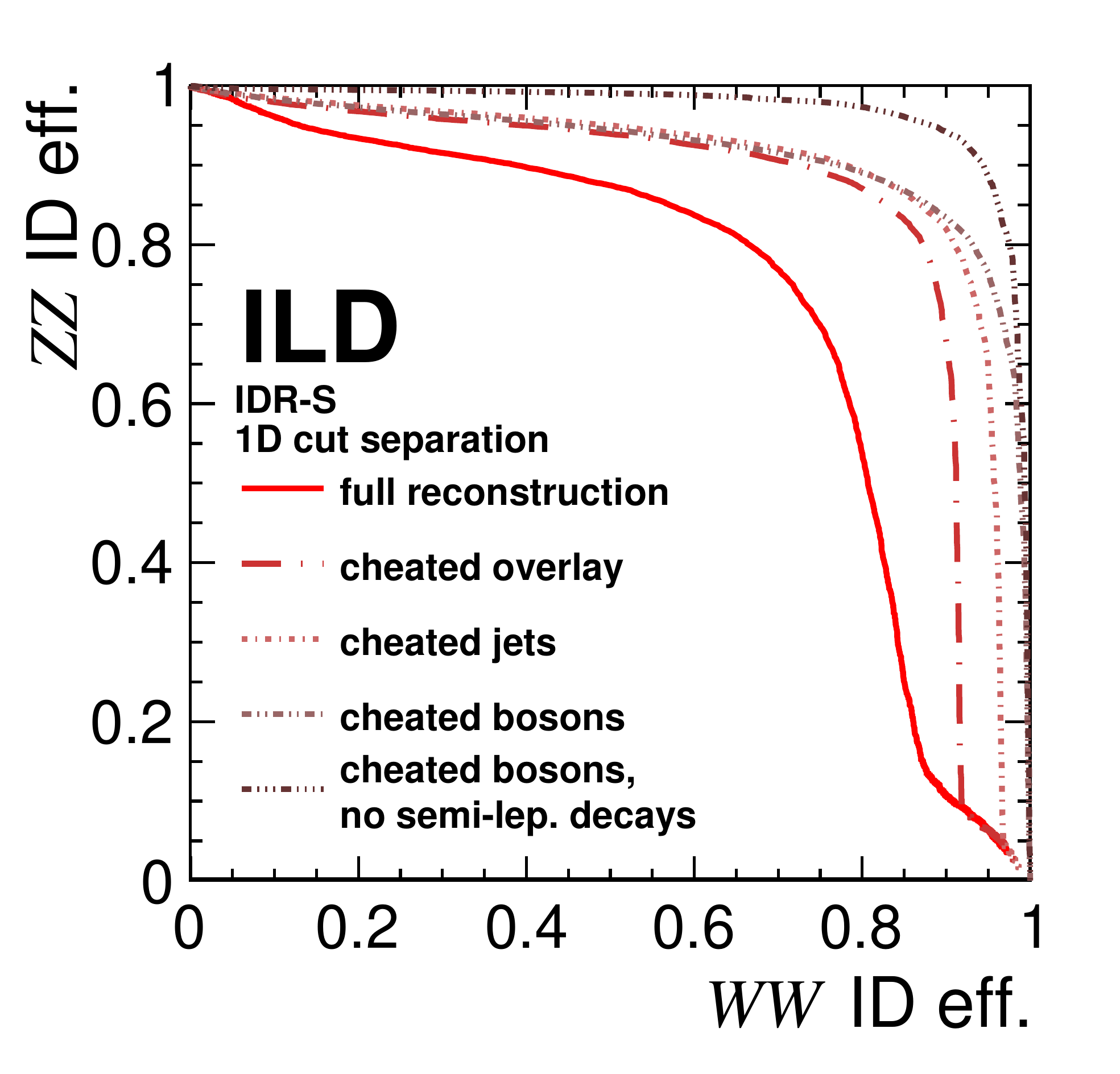}
    \caption{}
    \label{SUBFIG:IDRS_sep_curves_highQ2}
  \end{subfigure}
  
  \begin{subfigure}[t]{0.5\textwidth}
    \centering
    \includegraphics[width=\textwidth]{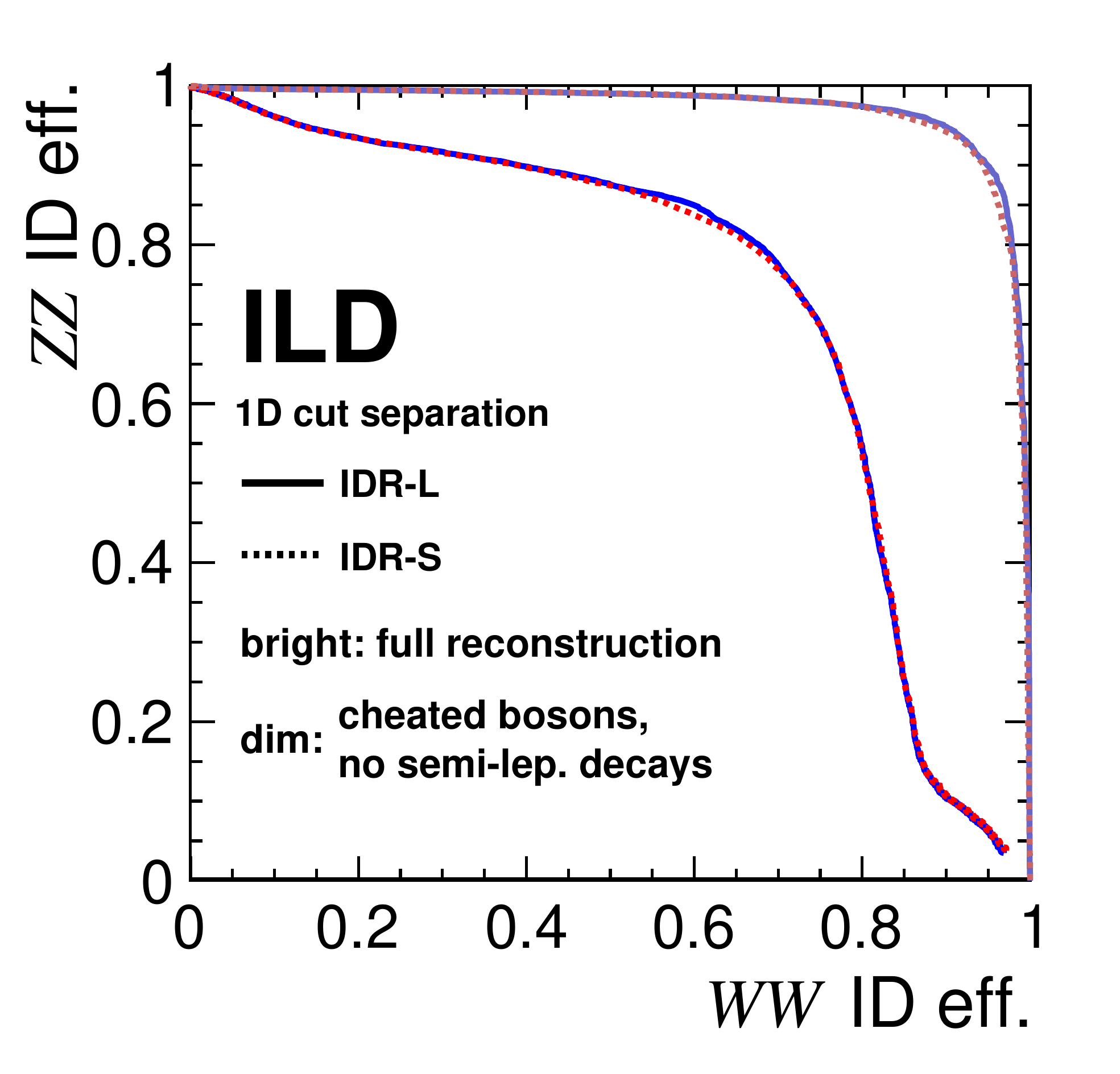}
    \caption{}
    \label{SUBFIG:IDR_sep_curves_ls_highQ2}
  \end{subfigure}
  \caption{
    Receiver operating characteristic (ROC) curve for a simple $WW/ZZ$ classification cut in the mass average distribution of the two reconstructed vector boson candidates.
    The dataset with a $m_{VV}>500\,$GeV restriction on generator level is used.
    In \subfigref{SUBFIG:IDRL_sep_curves_highQ2} and \subfigref{SUBFIG:IDRS_sep_curves_highQ2} for different levels of idealized reconstruction for the large and small ILD models, respectively.
    \subfigref{SUBFIG:IDR_sep_curves_ls_highQ2} comparing the large and small ILD models at the levels of full high level reconstruction and of fully idealized high level reconstruction.
  }
  \label{FIG:IDRSepCurves_highQ2}
\end{figure}

For completeness the mass distributions (fig.\ \ref{FIG:MassPlotsIDRL_highQ2}) and separation curves (fig.\ \ref{FIG:IDRSepCurves_highQ2}) for the dataset with a $m_{VV}>500\,$GeV restriction on generator level are provided here. 

\end{document}